\documentclass[5p,times, authoryear]{elsarticle}
\usepackage[utf8]{inputenc}
\usepackage{cancel}
\usepackage[dvipsnames]{xcolor}
\usepackage{amsmath}
\usepackage{url}
\usepackage{multirow}
\usepackage{xcolor}
\usepackage{algorithm}
\usepackage{algorithmic}
\usepackage{graphicx}
\usepackage{verbatim}
\graphicspath{ {./figure/} }
\usepackage{babel}
\addto\captionsenglish{}
\usepackage{xcolor}
\usepackage{tabularx}
\usepackage{amssymb}
\usepackage{amsthm}
\usepackage{array, boldline, makecell, booktabs}

\newtheorem{definition}{Definition}

\renewenvironment{proof}{{\bfseries Proof.}}{\qed}
\newtheorem{theorem}{Theorem}
\makeatletter
\def\zapcolorreset{\let\reset@color\relax\ignorespaces}
\def\colorrows#1{\noalign{\aftergroup\zapcolorreset#1}\ignorespaces}
\makeatother

\usepackage{mathtools}
\newcommand{\mathdash}{\relbar\mkern-19mu\relbar}

\usepackage{array,multirow}
\addto\captionsenglish{}
\usepackage{xfrac}
\usepackage{dblfloatfix}

\usepackage{etoolbox}
\makeatletter
\patchcmd{\@makecaption}
 {\scshape}
 {}
 {}
 {}
\makeatother

\usepackage{xcolor}
\usepackage{tabularx}
\usepackage{amssymb}
\usepackage{amsthm}
\usepackage{array, boldline, makecell, booktabs}
\renewenvironment{proof}{{\bfseries Proof.}}{\qed}
\usepackage{multirow}
\usepackage{enumitem}
\usepackage{diagbox}
\usepackage{afterpage}

\usepackage{color,soul}
\usepackage{array,multirow}
\usepackage{caption}
\newcommand{\specialcell}[2][c]{%
 \begin{tabular}[#1]{@{}c@{}}#2\end{tabular}}
\usepackage{microtype}
\usepackage{booktabs}
\usepackage[format=hang,font={small,bf}]{caption}
\usepackage[utf8]{inputenc}

\usepackage{color,soul}
\setul{0.5ex}{0.3ex}
\definecolor{Green}{rgb}{0,1,0}
\setulcolor{Green}
\usepackage{soul}
\usepackage[mathscr]{euscript}

\DeclareSymbolFont{rsfs}{U}{rsfs}{m}{n}
\DeclareSymbolFontAlphabet{\mathscrsfs}{rsfs}
\usepackage{lipsum}
\makeatletter
\def\ps@pprintTitle{%
 \let\@oddhead\@empty
 \let\@evenhead\@empty
 \def\@oddfoot{}%
 \let\@evenfoot\@oddfoot}
\makeatother

\begin{document}
\begin{frontmatter}

\title{Effective Graph-Neural-Network based Models for Discovering Structural Hole Spanners in Large-Scale and Diverse Networks}

\author[First]{Diksha Goel\corref{cor1}}
\ead{diksha.goel@adelaide.edu.au}

\author[Second]{Hong Shen}
\ead{shenh3@mail.sysu.edu.cn}

\author[Third]{Hui Tian}
\ead{hui.tian@griffith.edu.au}

\author[First]{Mingyu Guo}
\ead{mingyu.guo@adelaide.edu.au}

\cortext[cor1]{Corresponding author.}

\address[First]{School of Computer Science, University of Adelaide, Australia.}
\address[Second]{School of Computer Science and Engineering, Sun Yat-sen University, Guangzhou, China.}
\address[Third]{School of Information and Communication Technology, Griffith University, Australia.}

\begin{abstract}
A Structural Hole Spanner (SHS) is a set of nodes in a network that act as a bridge among different otherwise disconnected communities. Numerous solutions have been proposed to discover SHSs that generally require high run time on large-scale networks. Another challenge is discovering SHSs across different types of networks for which the traditional one-model-fit-all approach fails to capture the inter-graph difference, particularly in the case of diverse networks. Therefore, there is an urgent need of developing effective solutions for discovering SHSs in large-scale and diverse networks. Inspired by the recent advancement of graph neural network approaches on various graph problems, we propose graph neural network-based models to discover SHS nodes in large scale networks and diverse networks. We transform the problem into a learning problem and propose an efficient model GraphSHS, that exploits both the network structure and node features to discover SHS nodes in large scale networks, endeavouring to lessen the computational cost while maintaining high accuracy. To effectively discover SHSs across diverse networks, we propose another model Meta-GraphSHS based on meta-learning that learns generalizable knowledge from diverse training graphs (instead of directly learning the model) and utilizes the learned knowledge to create a customized model to identify SHSs in each new graph. We theoretically show that the depth of the proposed graph neural network model should be at least $\Omega(\sqrt{n}/\log n)$ to accurately calculate the SHSs discovery problem. We evaluate the performance of the proposed models through extensive experiments on synthetic and real-world datasets. Our experimental results show that GraphSHS discovers SHSs with high accuracy and is at least 167.1 times faster than the comparative methods on large-scale real-world datasets. In addition, Meta-GraphSHS effectively discovers SHSs across diverse synthetic networks with an accuracy of 96.2\%.
\end{abstract}

\end{frontmatter}

\section{INTRODUCTION}
The last decade witnessed tremendous growth of various large-scale networks, such as biological, semantic, collaboration, criminal and social networks. There is a huge demand for efficient and scalable solutions to study the properties of these large networks. A network consists of communities where the nodes share similar characteristics, and these communities are crucial for information diffusion in the network {\citep{chen2019contextual, goel2022discovering}}. The nodes having connections with the diverse communities get positional advantages in the network. This notion serves as a base for the \textit{Theory of} \textit{Structural Holes} \citep{burt2009structural}. The theory states that the \textit{\textbf{Structural Holes (SH)}} are the positions in the network that can bridge different communities and bring the beholders into an advantageous position. The absence of connections between different communities creates gaps, which is the primary reason for the formation of SHs in the network \citep{lou2013mining}. 

\begin{figure}[t!]
 \centering
 \includegraphics[width=0.65\columnwidth]{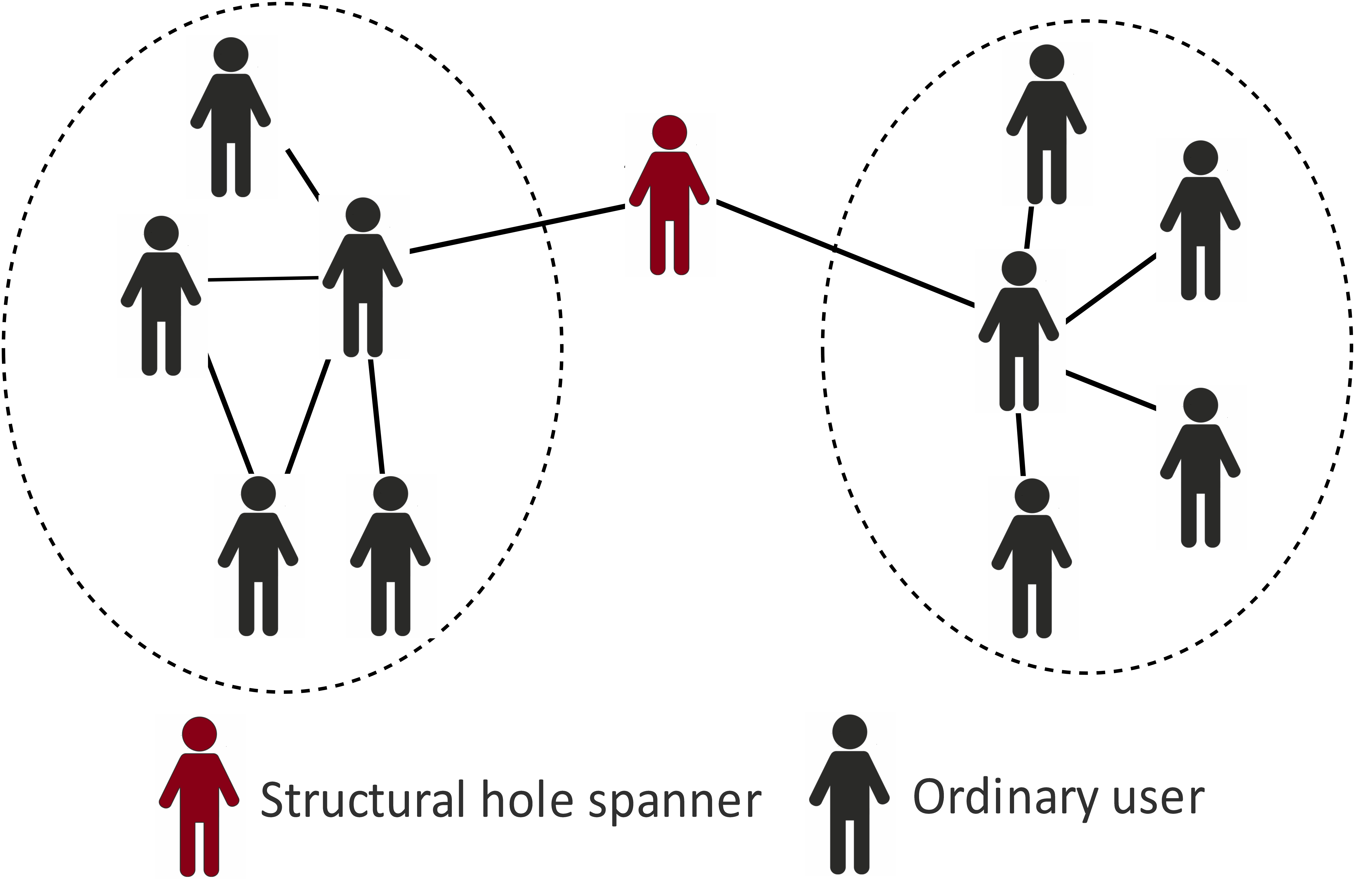}
 \caption{Illustration of SHS in the network.}
 \label{fig:SHS}
\end{figure}

The nodes that fill SHs by bridging different communities are known as \textit{\textbf{Structural Hole Spanners}} \citep{lou2013mining, goel2023discovering}. SHSs get various positional benefits such as access to novel ideas from diverse communities, more control over information flow etc. Figure \ref{fig:SHS} shows the SHS between communities in the network. There are many vital applications of SHSs, such as community detection, opinion control, information diffusion, viral marketing {\citep{gupta2020overlapping, kuhlman2013controlling, bonifazi2022approach, zhang2019most,  castiglione2020cognitive, jain2024comprehensive, abdulsatar2024towards}}, etc. In case of an epidemic disease, discovering SHSs and quarantining them can help stop the spread of infection. In addition, SHSs can be used to advertise a product to different groups of users for viral marketing.

\begin{figure*}[t!]
 \centering
 \includegraphics[width=0.6\paperwidth]{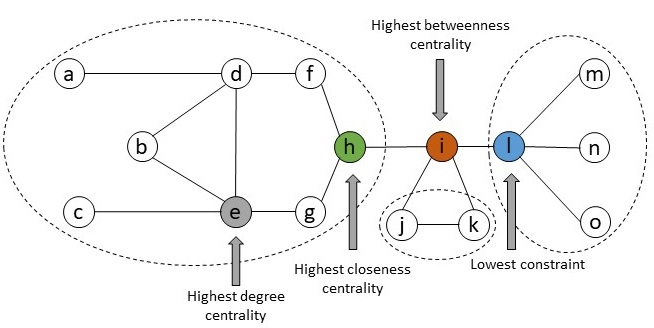}
 \caption{Comparison of various centrality measures.}
 \label{fig:cent}
\end{figure*}

A number of centrality measures such as Closeness Centrality \citep{rezvani2015identifying}, Constraint \citep{burt1992structural}, Betweenness Centrality (BC) \citep{freeman1977set} exist in the literature to define SHSs. SHS nodes lie on the maximum number of shortest paths between the communities \citep{rezvani2015identifying}; removal of the SHS nodes will disconnect multiple communities and block information flow among the nodes of the communities \citep{lou2013mining}. Based on this, we have two implications about the properties of SHSs; 1) SHSs bridge multiple communities; 2) SHSs control information diffusion in the network. Figure \ref{fig:cent} illustrates the comparison of various centrality measures in a network. The figure shows that node $i$ holds a vital position in network, and the shortest path between the nodes of three communities passing-through node $i$, and removing node $i$ will block the information propagation between the nodes of these communities. In contrast, the impact of removal of other nodes is comparatively less significant. Since removal of a node with the highest betweenness centrality disconnects maximum number of communities and blocks information propagation between the nodes of the communities, therefore, we adopt the betweenness centrality measure for defining SHSs in the network. Goyal et al. \citep{goyal2007structural} defined the node that lies on a large number of shortest paths as SHS, which is similar to the betweenness centrality. BC quantifies a node's control on the information flow in the network and discovers those nodes that act as a bridge between different communities. 
Brandes algorithm  is the best-known method for calculating the BC scores of the nodes and has a run time of $\mathcal{O}(nm)$ {\citep{brandes2001faster}}. \\

\noindent \textbf{Challenges:} Several studies have been conducted for discovering SHSs in the network {\citep{lou2013mining, he2016joint, xu2019identifying, li2019distributed}}. Lou et al. \citep{lou2013mining} developed an algorithm for finding SHSs by assuming that community information is given in advance. However, discovering communities in a large network is a challenging task. He et al. \citep{he2016joint} designed a harmonic modularity solution
that discovers both SHSs and communities in the network. The authors assume that every node belongs to one community only, but a node may belong to many communities in the real world. Although there are numerous solutions that address the SHSs identification problem; however, there are still \textit{{challenges}} that need to be addressed, such as:
\vspace{-.06in}
\begin{enumerate}[leftmargin=0.5cm]
    \item \textbf{Discovering SHS nodes efficiently in large scale networks:} For small networks, we can discover SHSs by computing the BC score of the nodes using Brandes algorithm; however, for large networks, Brandes algorithm’s run time of $O(nm)$ is very high {\citep{brandes2001faster}}. Therefore, we need efficient solutions for discovering SHSs in large scale networks.
    \vspace{-.06in}
    \item \textbf{Discovering SHS nodes effectively in diverse networks:} For discovering SHSs in different types of network, traditional learning techniques fail to work because their one-model-fit-all approach neglect the inter-graph differences, especially when the graphs belong to diverse domains. Besides, re-training the model again on different types of large networks is a time-consuming process. Therefore, it is crucial to have a model which is aware of differences across the graphs and customizes accordingly, avoiding the requirement of re-training the model on every type of network individually.
\end{enumerate}

To address the challenges mentioned above and inspired by the recent advancements of Graph Neural Network (GNN), we propose message-passing GNN based models to discover SHS nodes. GNNs  are Neural Network architectures designed for graph structured data {\citep{thekumparampil2018attention, kipf2016semi}}. GNNs are used as graph representation learning models and learn node representations by aggregating feature information from the local graph neighbourhood {\citep{joshi2019efficient, djenouri2022hybrid}}. GNNs have shown exceptional results on various graph mining problems \citep{horta2021extracting, ji2021temporal}; therefore, we investigate the power of GNNs for solving SHS identification problem.

\textit{In this paper, we aim to discover SHS nodes in large-scale networks, endeavouring to reduce the computational cost while maintaining high accuracy, and in different types of networks effectively without the need of re-training the model on individual network datasets to adapt to cross-network property changes.} 

We transform the SHS discovery problem into a \textit{learning problem} and propose two GNN based models, GraphSHS and Meta-GraphSHS. In order to address the first challenge mentioned above, we propose \textbf{\textit{GraphSHS}} \textit{\textbf{(\underline{Graph} neural network for \underline{S}tructural \underline{H}ole \underline{S}panners)}}, a graph neural network-based model for efficiently discovering SHSs in large scale networks. GraphSHS exploits both the network structure and features of nodes to learn the low-dimensional node embeddings. 
In addition, unlike traditional Deep Learning approaches that assume a transductive setting, GraphSHS assumes an inductive setting. GraphSHS is generalizable to new nodes of the same graph or even to the new graphs from the same network domain. 
Our experimental results demonstrate that the idea of designing graph neural network based model to discover SHSs in large scale networks provides a \textit{\textbf{significant run time advantage}} over other algorithms. Apart from the run time efficiency, GraphSHS achieves competitive or better accuracy in most of the cases than the baseline algorithms. 

To address the second challenge, we propose \textbf{\textit{Meta-GraphSHS}} \textit{\textbf{(\underline{Meta-}learning based \underline{Graph} neural network for \underline{S}tructural \underline{H}ole \underline{S}panners)}} to effectively discover SHSs across diverse networks. In the case of diverse graphs, there exist inter-graph differences due to which GraphSHS can not effectively discover SHSs across diverse networks. Therefore, instead of directly learning the model, we learn the generalizable knowledge (parameters) from diverse training graphs and utilize the learned knowledge to create a customized model by fine-tuning the parameters according to each new graph\footnote{The generalizable knowledge act as a good initialization point (good set of parameters) for the new customized model. The generalized parameters are fine-tuned using the labelled nodes of new unseen graphs.}. Meta-GraphSHS uses meta-learning to learn generalizable parameters from the training graphs that are different from the testing graphs we are considering, and the goal is to reach an \textbf{\textit{``almost trained"}} model that can be quickly adapted to create a customized model for the new graph under consideration within a few gradient steps. The goal of Meta-GraphSHS is to observe many graphs from different domains and use the learned knowledge to identify SHS on any new graphs,\textit{ enabling quick adaptation and higher accuracy}.

Once our proposed model is trained, it can be applied repeatedly for future arriving data; therefore, we consider primarily the run time of applying the model and regard the training process is done offline, as the common practice in machine learning literature.

\textit{Our experimental results show that both the proposed graph neural network models GraphSHS and Meta-GraphSHS are highly efficient and effective in discovering SHSs in large scale networks and diverse networks, respectively. We evaluate the performance of GraphSHS on synthetic datasets, and the results show that GraphSHS is at least 58 times faster than baselines and achieves higher or competitive accuracy than baselines. In addition, GraphSHS is at least 167.1 times faster than the baselines on real-world networks, illustrating the efficiency advantage of the proposed GraphSHS model on large-scale networks. We evaluate the performance of Meta-GraphSHS on a diverse set of synthetic and real-world graphs, and the results show that Meta-GraphSHS identifies SHSs with high accuracy, i.e., 96.2\% on synthetic graphs and outperforms GraphSHS by 2.7\% accuracy, demonstrating the importance of designing separate model for discovering SHSs in diverse networks. Additionally, we also conduct parameter sensitivity analysis to analyze the impact of parameters on the performance of proposed models. In order to determine the applicability of the proposed model GraphSHS in the dynamic network, we perform experiments on synthetic graphs and found that our model is at least 89.8 times faster than the existing baseline.}

\noindent The contributions of the paper are summarized below:
\begin{itemize}[leftmargin=*,noitemsep,topsep=0pt]
\item \textbf{GraphSHS model.} We propose an efficient graph neural network-based model GraphSHS that discovers SHSs in large scale networks and achieves considerable efficiency advantage while maintaining high accuracy compared to existing baselines. 
\item \textbf{Meta-GraphSHS model.} We propose another model Meta-GraphSHS that combines meta-learning with graph neural network to discover SHS nodes across diverse networks effectively. This model learns a generalized knowledge from diverse graphs that can be utilized to create a customized inductive model for each new graph, in turn avoiding the requirement of repeated model training on every type of diverse graph.

\item \textbf{Inductive setting.} We use an inductive setting, where our GraphSHS model is generalizable to new nodes of the same graph or even to the new graphs from the same network. In addition, the proposed Meta-GraphSHS model is generalizable to unseen graphs from diverse networks.

\item \textbf{Theoretical analysis.} We theoretically show that our message-passing architecture of GraphSHS is sufficient to solve the SHSs identification problem under sufficient conditions on its node attributes, expressiveness of layer, architecture's depth and width. In addition, we show that the depth of the model should be at least  $\Omega(\sqrt{n}/\log n)$ to accurately solve the SHSs identification problem.

\begin{table*}[!t] 
\caption{Summary of SHSs identification solutions.}
\label{related work}
\renewcommand{\arraystretch}{1}
\centering 
\begin{tabular}{>{}p{1.7cm}>{}p{2cm}cp{4cm}>{}p{2.5cm}>{}p{2.5cm}} \hlineB{1.5} 
\textbf{Category} &\textbf{Reference} & \textbf{Method} & \textbf{Main idea} & \textbf{Pros} & \textbf{Cons} \\ \hlineB{1.5}

& \multirow{2}{*}{ \citep{lou2013mining}} & \specialcell[t]{HIS\\MaxD} & SHS connects opinion leaders of the various communities & Proved convergence of model & Require prior community information \\ \cmidrule{2-6}
& \multirow{2}{*}{\citep{he2016joint}} & \multirow{2}{*}{HAM} & The authors used harmonic function to identify SHSs &  Jointly discover SHSs and communities & High computational cost\\ \cmidrule{2-6}
{Information Propagation} & \multirow{2}{*}{\citep{xu2019identifying}} & \specialcell[t]{maxBlock\\maxBlockFast} & SHSs are likely to connects multiple communities and have strong relations with these communities & Less computational cost & Does not work for diverse networks\\ \cmidrule{2-6}

& \multirow{2}{*}{ \citep{li2019distributed}} & \multirow{2}{*}{ESH} & The authors designed entropy-based mechanism that uses distributed parallel computing & Less computational cost & Does not work for diverse networks \\ \hline

& \multirow{2}{*}{\citep{tang2012inferring}} & \specialcell[t]{2-step\\algorithm} & The model considers the shortest path of length two that pass through the node & Does not require community information & It fails to work in case a node is densely linked to many communities \\ \cmidrule{2-6}
 & \multirow{2}{*}{ \citep{rezvani2015identifying}} & \specialcell[t]{ICC\\BICC\\AP\_BICC} & Eliminating SHSs from the network leads to an increase in average shortest distance of the network & Only used topological network structure & Does not work for diverse networks \\\cmidrule{2-6}
Network Centrality & \multirow{2}{*}{\citep{xu2017efficient}} & \specialcell[t]{Greedy\\AP\_Greedy} & The authors used inverse closeness centrality to discover SHSs & Does not require community information & Does not work for diverse networks\\ \cmidrule{2-6}
 
& \multirow{2}{*}{\citep{ding2016method}} & \multirow{2}{*}{V-Constraint} & The authors used ego-network of the node to discover SHSs & Detects key nodes occupying SHs in network & Ego network may not capture the global importance of the node \\ \cmidrule{2-6}
 
& \multirow{2}{*}{\citep{goel2021maintenance}} & \specialcell[t]{Decremental\\algorithm} & The authors used total pairwise connectivity metric to discover SHSs & Discover SHSs in dynamic networks & Doesnot work for incremental updates\\ \cmidrule{2-6}
& \multirow{2}{*}{\citep{zhang2020finding}} & \multirow{2}{*}{FSBCDM} & The author used community forest-based model utility to discover SHSs & Jointly discover SHSs and communities & Higher computational complexity \\\hline

Machine Learning & \multirow{2}{*}{\citep{gong2019identifying}} & \specialcell[t]{Machine learning\\model} & The authors used various cross-site and ego network features of the nodes & Achieves high accuracy & Depend on HIS \citep{lou2013mining} to obtain
ground-truth \\\hlineB{1.5}

\end{tabular} 
\end{table*}

\item \textbf{Extensive experiments.} We conduct extensive experiments on synthetic networks and real-world networks of varying scales. The results show that the proposed model GraphSHS is at least 167.1 times faster than the baselines on real-world networks and at least 58 times faster on synthetic networks. In addition, Meta-GraphSHS discovers SHSs across diverse networks with an accuracy of 96.2\%.
\end{itemize}

\noindent \textbf{Organization.} Section 2 reviews the work done by the researchers in the field. Section 3 discusses the preliminaries and problem definition. Section 4 discusses the details of the proposed models GraphSHS and Meta-GraphSHS. Section 5 reports and discusses the experimental results. Finally, Section 6 concludes the paper and present future work.

\section{RELATED WORK}
The theory of SH \citep{burt2009structural} was initially introduced by Burt to discover the important individuals of the organization and was further investigated by \citep{ahuja2000collaboration, burt2007secondhand}. {There are numerous pioneering works for discovering SHSs, and the work can be classified as information propagation-based solutions, centrality-based solutions, and machine learning-based solutions \citep{goel2023enhancing}. In order to provide a comprehensive overview of these approaches, Table  \ref{related work} presents the summary of SHS identification solutions. In the following section, we delve into the state-of-the-art solutions for identifying SHSs. Subsequently, we also explore Meta-Learning based approaches that address similar research problems.}

\subsection{Information Propagation based Solutions}
The solutions based on information propagation aim to identify the SHS nodes that either maximize the information flow or whose removal maximally disrupts the information flow in the network. Lou et al. \citep{lou2013mining} designed an algorithm for discovering SHSs in the network, assuming that community information is given in advance. However, the solution fails to work in case community information is not known in advance. He et al. \citep{he2016joint} designed a Harmonic Modularity (HAM) solution that simultaneously discovers SHSs and communities in the network. The authors investigated the interaction type among the nodes, mainly for bridging nodes, to differentiate the SHS nodes from the normal ones. The algorithm assumes that every node belongs to one community only, but a node can be linked to many communities in the real world. Motivated by \citep{burt2001closure, burt2011structural}, Xu et al. \citep{xu2019identifying} designed a fast algorithm to detect SHS that connects multiple communities and has strong relations with these communities. The authors argued that eliminating spanners results in blocking maximum information in the network. Li et al. \citep{li2019distributed} designed a model for discovering SHSs using distributed and parallel processing. Unlike other techniques, the authors introduced an entropy-based mechanism and applied distributed computing. Zhang et al. \citep{zhang2016identifying} proposed a vote rank algorithm to identify top-$k$ decentralized spreaders with the best spreading ability. This algorithm uses a voting scheme to rank nodes in a graph, where each node votes for its in-neighbors, and the node with the highest number of votes is selected in each iteration. 

\subsection{Network Centrality based Solutions} 
The solutions based on network centrality aim to discover the nodes located at advantageous positions. Tang et al. \citep{tang2012inferring} proposed a two-step mechanism for discovering SHSs. For each node, the authors only considered the shortest path of length two while ignoring others. Rezvani et al. \citep{rezvani2015identifying} designed a solution for discovering SHS nodes based on inverse closeness centrality. The authors argued that removing SHSs from the network results in an increase in the shortest distance of the network. They further improved the solution and proposed a bounded inverse closeness centrality solution. Motivated by \citep{rezvani2015identifying}, Xu et al. \citep{xu2017efficient} proposed an efficient solution for discovering SHSs. The solution does not require any community information. In addition, the authors claim that their algorithm is able to capture the features of SHSs with high accuracy. Ding et al. \citep{ding2016method} proposed V-Constraint for discovering SHSs. The author used node features such as the degree of the node and various other topological features of the neighbors. Goel et al. \citep{goel2021maintenance} designed a decremental algorithm for discovering SHSs in dynamic networks. The authors reuse the previous knowledge to discover SHSs for the current network to avoid unnecessary recomputations. Zhang et al. \citep{zhang2020finding} designed a community forest model to detect SHSs. The authors argued that local features might not be suitable for discovering spanner nodes in the network. {Maier et al. \citep{maier2021saturated} proposed a centrality measure in a graph called saturated BC sets, to identify a group of nodes that exhibit control over information flow in a graph. The authors defined saturated BC sets as the set of nodes with maximal BC than any other set of nodes. The authors proposed an algorithm to detect sets of nodes that exhibit high control over information flow while minimizing set size. The author aims to find a set of nodes such that their collective BC is greater than the BC of any proper subset and is smaller or equal to the BC of any possible superset. In contrast, we aim to find the top-$k$ node with the highest BC, and hence we focus on the individual node’s BC score rather than the top-$k$ node’s collective BC score.}

\subsection{Machine Learning based Solutions} 
Machine learning-based solutions use the features of the nodes to identify SHSs. Gong et al. \citep{gong2019identifying} proposed a supervised learning solution to detect SHSs. The authors used a location-based social network, and their model relies on users' demographic information and statistics of user-generated content from Twitter. In contrast, our proposed model GraphSHS is designed to work using structural features of the network. {Luo et al. \cite {luo2022bridge} proposed a deep learning model for detecting bridge nodes in the networks. The model considers multi-dimensional attributes and structural characteristics of nodes. The authors used existing graph neural network models to process the input graph, and added fully connected layers to improve the model classification accuracy. The model presented in the paper utilizes the Louvain algorithm to determine community structure, which is then utilized to label bridge nodes using three algorithms: gateway local rank, common centrality index, and neighbour-based bridge node centrality. Notably, our proposed model GraphSHS does not rely on the community structure to identify SHS nodes in the network, which is an advantage over \cite {luo2022bridge}.

{Luo et al. \citep{luo2020detecting} introduced ComSHAE, a deep learning-based approach that utilizes a graph convolutional neural network-based Auto-Encoder to simultaneously identify communities and SHSs in networks. However, their evaluations were limited to small-scale networks. In contrast, our proposed model demonstrates the capability to scale to larger networks and discover SHSs across a variety of network types. Hamilton et al. \citep{hamilton2017inductive} presented GraphSAGE, a method that leverages node features and neighborhood sampling to generate embeddings for both seen and unseen nodes, effectively addressing the limitations of transductive methods. Velickovic et al. \citep{velivckovic2017graph} introduced Graph Attention Networks (GAT), which enables nodes to assign varying weights to their neighbors' features without relying on costly matrix operations or prior knowledge of the graph structure. Xu et al. \citep{xu2018powerful} introduced a highly expressive GNN architecture known as Graph Isomorphism Network (GIN), which is on par with the Weisfeiler-Lehman graph isomorphism test. However, it's worth noting that GraphSAGE and GAT are primarily designed to discover nodes in a single type of network. In contrast, one of our proposed model focuses on discovering SHS nodes across diverse networks.}

\subsection{{Meta Learning based Solutions}} 

{We explored various machine-learning methods and discovered that Meta-Learning techniques can be effectively used to identify the SHS nodes in diverse networks. Meta-Learning design models that can learn to learn and are able to adapt to new tasks very quickly. Wen et al. \citep{wen2021meta} presented MI-GNN, a meta-inductive framework for customized inductive node classification across graphs. Ding et al. \citep{ding2021few} introduced Graph Deviation Networks for few-shot network anomaly detection, using labeled anomalies and cross-network meta-learning. Liu et al. \citep{liu2022few} proposed Meta-GPS for few-shot node classification on attributed networks, achieving significant performance improvements with expressive node representations. Zhou \citep{zhou2019meta} proposed a meta-learning framework that enhances few-shot node classification on non-Euclidean graphs. Huang et al. \citep{huang2020graph} developed meta-learning algorithm for graphs that leverages local subgraphs to transfer subgraph-specific information, enabling fast adaptation to new tasks with limited data. Liu et al. \citep{liu2019learning} proposed Gated Propagation Network, a meta-learning approach that explicitly relates tasks through a graph describing output dimension relations, improving few-shot learning by propagating messages between class prototypes.}

Our proposed model not only learns from the node's features but also from the network structure. Unlike traditional learning methods, our model follows an inductive setting and is generalizable to unseen graphs. Furthermore, utilizing meta-learning enables us to discover SHS nodes from different networks. Scalability is another issue due to which some of the existing solutions \citep{lou2013mining, he2016joint} do not work for large-scale networks. However, our model is scalable to large networks and exhibits a significant advantage in run time over other algorithms. SHS identification solution proposed by Lou et al. \citep{lou2013mining} requires community information; however, our solution does not require prior community information. Besides, our proposed solution performs better than \citep{goel2021maintenance} for dynamic networks.

\begin{table}[t!] 
\caption{List of Abbreviations}
\label{abbreviations}
\renewcommand{\arraystretch}{1}
\centering 
\begin{tabular}{>{}p{2cm}>{}p{5.5cm}} \hlineB{1.5} 
\textbf{Abbreviation} & \textbf{Full Form} \\ \hlineB{1.5}
SH & Structural Hole \\ 
SHS & Structural Hole Spanner \\ 
BC & Betweenness Centrality \\
GNN & Graph Neural Network \\ 
ER & Erdos Renyi \\
SF &  Scale Free \\
CC & Closeness Centrality \\ 
SP & Shortest Path \\ 
GraphSHS & Graph neural network for Structural Hole Spanners\\
Meta-GraphSHS & Meta-learning based Graph neural network for Structural Hole Spanners \\
\hlineB{1.5}
\end{tabular} 
\end{table}

\section{PRELIMINARIES AND PROBLEM DEFINITIONS}
\subsection{Preliminaries}
\noindent \textbf{\textit{Notations.}} A network can be represented as an undirected graph $G = (V, E)$, where $V$ is the set of nodes (users), and $E$ is the set of edges (the relationship between users). Let $n=|V|$ and $m=|E|$. We use $\vec{x}(i)$ to represent the feature vector of node $i$ and $h^{(l)}(i)$  to denote the embedding of node $i$ at the $l^{th}$ layer of the model, where $l = (1,2,...,L)$. The neighbors of node $i$ are represented by $N(i)$, and the degree of node $i$ is represented by $d(i)$. {Table \ref{abbreviations} presents the list of abbreviations and Table\ref{notations} presents the list of symbols used in this paper.} \\

\noindent \textbf{\textit{Graph Neural Networks.}}
Graph Neural Networks (GNNs) are designed by extending Deep Learning approaches for the graph-structured data and are used in diverse fields, including computer vision, graph problems etc. GNNs are used to learn the graph data representations. Motivated by the success of Convolution Neural Network, various Graph Neural Network architectures are designed. One such architecture is Graph Convolutional Network, which uses an aggregation mechanism similar to the mean pooling {\citep{kipf2016semi}}. Graph Attention Network is another Graph Neural Network architecture that uses an attention mechanism for aggregating features from the neighbors {\citep{velivckovic2017graph}}. Existing GNN architectures mostly follow message-passing mechanism. These GNNs execute graph convolution by aggregating features from neighbours, and stacking many layers of GNN to capture far-off node dependencies.\\

\begin{table}[t!] 
\caption{Table of symbols}
\label{notations}
\renewcommand{\arraystretch}{1.2}
\centering 
\begin{tabular}{ll} \hlineB{1.5} 
\textbf{Symbol} & \textbf{Definition} \\ \hlineB{1.5}
$G$ & Original graph \\ 
$V,E$ & Set of nodes and edges \\ 
$n,m$ & Number of nodes and edges\\
$k$ & Number of SHSs \\
$l$ & Index of aggregation layer \\ 
$L$ & Total number of aggregation layers \\
$||$ & Concatenation operator \\
$\sigma$ & Non-linearity \\ 
$z(i)$ & Final embedding of node $i$ \\
$y(i)$ & Label of node $i$ \\
$\vec{x}(i)$ & Feature vector of node $i$\\ 
$d(i)$ & Degree of node $i$ \\
$N(i)$ & Neighbors of node $i$ \\
$h^{(l)}(i)$ & Embedding of node $i$ at the $l^{th}$ layer \\ 
\hlineB{1.5}
\end{tabular} 
\end{table}

\begin{figure}[h!]
 \centering
 \includegraphics[width=0.8\columnwidth]{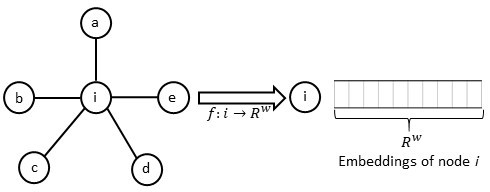}
 \caption{Network Embedding: embedding of node $i$ in $R^w$ space.}
 \label{fig:embed}
\end{figure}

\noindent \textbf{\textit{Network Embedding.}} Network embedding is a mechanism that maps the nodes of the network to a low-dimensional vector representation {\citep{cui2018survey}}. It aims to encode the nodes in such a way that the resemblance in the embedding space approximates the resemblance in the network {\citep{aguilar2021novel}}. These embeddings can then be utilized for various graph problems such as classification, regression etc. Figure \ref{fig:embed} illustrates an example of node embedding.\\

\noindent \textbf{\textit{Meta-Learning.}}
Meta-Learning 
aims to learn efficiently and generalize the learned knowledge to the new tasks. There are various meta-learning approaches such as black-box methods, gradient-based methods and non-parametric learning methods {\citep{andrychowicz2016learning, finn2017model, chen2019closer}}. Meta-Learning assumes that the prior learned knowledge is transferable among the tasks. The model trained on the training tasks can be adjusted to the new task using a small amount of labelled data or in the absence of any supervised knowledge. Meta-learning significantly improves the performance of the tasks that suffers from data deficiency problem. It learns the shared learning from the various tasks and adapts this knowledge to the unseen tasks, speeding up the learning process on new tasks.

\begin{definition}
\noindent \textbf{Betweenness Centrality.} The betweenness centrality $BC(v)$ of a node $v \in V$ is defined as \citep{freeman1977set}:
\begin{equation}
\begin{aligned}
BC(v) = \sum_{\substack{s\neq v\neq t\\v \in V}}{\frac{\sigma_{st}{(v)}}{\sigma_{st}}}
\end{aligned}
\end{equation}
\end{definition}
\noindent where $\sigma_{st}$ denotes the total number of shortest paths from node $s$ to $t$ and $\sigma_{st}{(v)}$ denotes the number of shortest paths from node $s$ to $t$ that pass through node $v$. We will use the term SHS score of a node and BC of a node interchangeably. \textit{We label $k$ nodes with the highest BC in the graph as \textit{Structural Hole Spanner nodes} and the rest as normal nodes.}

\subsection{Problem Definition}
In theory, the computation of Betweenness Centrality (discovering SHSs) is tractable as polynomial-time solutions exist; however, in practice, the solutions are computationally expensive. Currently, Brandes algorithm  is the best-known technique for calculating the BC of the nodes with a run time of $\mathcal{O}(nm)$ {\citep{brandes2001faster}}. However, this run time is not practically applicable, considering that even mid-size networks may have tens of thousands of edges. Computing the exact BC for a large scale network is not practically possible with traditional algorithms; consequently, we convert the SHS identification problem into a learning problem and then solve the problem. We formally define both the structural hole spanner discovering problems as follows:\\

\noindent \textbf{{Problem 1:}} \textbf{Discover SHS nodes in large scale networks.}

\noindent {\textbf{\textit{Input:}}} Training graph $G_{train}$, features and labels\footnote{Label of a node can either be SHS or normal.} of nodes in $G_{train}$, and test graph $G_{test}$.

\noindent \textbf{\textit{Goal:}} Design an inductive model GraphSHS (by training the model on $G_{train}$) to discover SHSs in new unseen large scale graph $G_{test}$. GraphSHS aims to achieve a considerable efficiency advantage while maintaining high accuracy. \\

\noindent \textbf{{Problem 2:}} \textbf{Discover SHS nodes in diverse networks.}

\noindent {\textbf{\textit{Input:}}} A set of training graphs $G_{train}=\{G_1, G_2, . . . , G_M\}$ from diverse domains, features and labels of nodes in $G_{train}$ and test graph $G_{test}$ in which the nodes are partially labeled.

\noindent \textbf{\textit{Goal:}} Design a model Meta-GraphSHS to discover SHS nodes across diverse networks effectively by learning generalized knowledge from diverse training graphs $G_{train}$. The generalized knowledge (parameters) is fine-tuned using labelled nodes from $G_{test}$ in order to obtain updated parameters that can be used to discover SHSs in $G_{test}$.\\

We address the above-discussed two problems by transforming them into learning problems and proposing two message-passing GNN-based models. Once the models are trained, the inductive setting of the models enables them to discover SHS nodes. The identified $k$ SHSs are the nodes with the highest SHS score (BC) in the network.

\section{PROPOSED MODEL}
This section discusses the proposed models GraphSHS and Meta-GraphSHS for discovering SHSs. We first discuss the network features that we extracted to characterize each node. We then discuss the proposed models GraphSHS and Meta-GraphSHS in detail.

\begin{figure*}[t!]
 \centering
 \includegraphics[width=0.8\paperwidth]{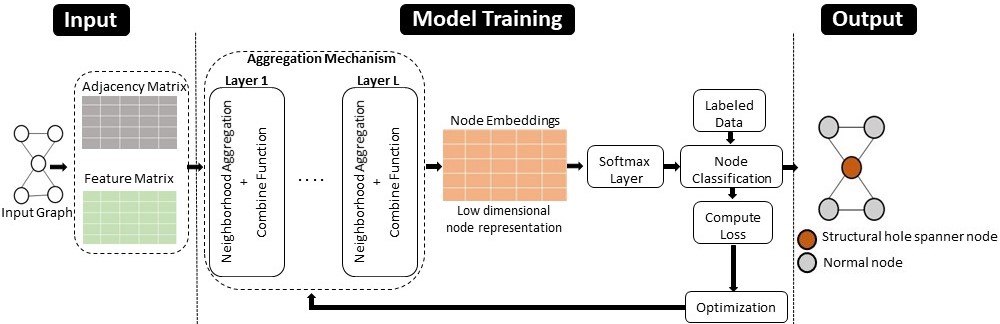}
 \caption{The overall architecture of the proposed model GraphSHS.}
 \label{fig:gnn_arch}
\end{figure*}

\subsection{Network Features} \label{features}
\begin{definition}

\textbf{\textit{r}-ego network}. The \textit{r}-ego network of a node $v \in V$ is the subgraph induced from $N_{r}(v)$ where $N_{r}(v) = \{u :dist_{uv}^{G} \leq r \}$ is $v's$ $r$-hop neighbors and $dist_{uv}^{G}$ denotes the distance between node $u$ and $v$ in graph $G$.
\end{definition}

\noindent We use three network features; effective size, efficiency and degree computed from the one-hop ego network of each node to characterize the node.\\

\noindent \textit{\textbf{Effective Size.}} The effective size is a measure of non-redundant neighbors of a node \citep{burt1992structural}. Effective size determines the extent to which neighbor $j$ is redundant with the other neighbors of node $i$. \\

\noindent \textit{\textbf{Efficiency.}} The efficiency is the ratio of the effective size of ego network of the node to its actual size \citep{burt1992structural}.  \\

\noindent \textit{\textbf{Degree.}} The degree of a node is the number of connections it has with the other nodes of the network.

\subsection{GraphSHS: Discovering SHSs in Large-Scale Networks}

In this section, we discuss our proposed message-passing graph neural network-based model \textbf{\textit{GraphSHS}} that aims to discover SHS nodes in large scale networks. 
Figure \ref{fig:gnn_arch}  illustrates the overall architecture of the proposed model GraphSHS. To discover SHSs, GraphSHS first maps each node to an embedding vector (low dimensional node representation) using the aggregation mechanism. GraphSHS then uses the embedding vector of each node to determine the labels of the node. The aggregation mechanism and the training procedure of GraphSHS are discussed below.

\subsubsection{Aggregation Mechanism} 
Our proposed aggregation mechanism computes the low dimensional node embeddings in two phases: 1) Neighborhood aggregation phase, where a node aggregates embeddings from its neighbors; 2) Combine function phase, where a node combines its own embedding to the aggregated neighbors embeddings. The procedure for generating embeddings of the nodes is presented in Algorithm \ref{gnn-algo}.\\

\noindent \textbf{Neighborhood Aggregation.}
For generating the node embeddings, GraphSHS first performs neighborhood aggregation by capturing feature information (embeddings) from the neighbors of the node. This process is similar to the \textit{message passing mechanism} of GNNs. Due to the distinctive properties exhibited by the SHS node (i.e., the SHS node act as a bridge, and its removal disconnects the network), we aggregate embeddings from all one-hop neighbors of the node. We describe the neighborhood aggregation as a weighted sum of embedding vectors and is given by:
\begin{equation}
\label{agg}
h^{(l)}{(N(i))} = \sum_{j\in N(i)}{\frac{h^{(l-1)}{(j)}}{d(i)}}
\end{equation}

\noindent where $h^{(l)}{(N(i))}$ denotes the embedding vectors aggregated from the neighbors $N(i)$ of node $i$ at the $l^{th}$ layer. The aggregated embedding from the neighbors of node $i$ is used to update node $i$'s embeddings. During the aggregation process, we utilize the degree $d$ of the node as a weight. We use the features of the nodes (as discussed in Section \ref{features}) to compute the initial embedding $h^{(0)}$ of the nodes. Let $\vec{x}(i)$ represents the feature vector of node $i$; GraphSHS initialize the initial embedding of node $i$ as:
\begin{equation}
h^{(0)}(i) = \vec{x}(i) \\
\end{equation}
Therefore, given a network structure and initial node features, neighborhood aggregation phase computes the embedding of each node by aggregating features from the neighbors of the nodes.\\

\begin{algorithm}[h!]
\caption{Generating node embedding using GraphSHS}
 \label{gnn-algo}
 \begin{algorithmic}[1]
 \renewcommand{\algorithmicrequire}{\textbf{Input:}}
 \renewcommand{\algorithmicensure}{\textbf{Output:}}
 \REQUIRE Graph: $G(V,E)$, Input features: $\vec{x}(i),\,\, \forall i \in V$, Depth: $L$, Weight matrices: $W^{(l)},\,\, \forall l \in \{1,..,L\}$, Non-linearity: $\sigma$
 \ENSURE Node embedding: $z{(i)}, \,\, \forall i \in V$ 
 \STATE $h^{(0)}(i) \leftarrow \vec{x}(i),\,\,\forall i \in V$ 
 \FOR{$l = 1$ to $L$}
 \FOR{$i \in V$}
 \STATE Compute $h^{(l)}{(N(i))}$ using Equation \ref{agg}
 \STATE Compute $h^{(l)}{(i)}$ using Equation \ref{comb}
 \ENDFOR
 \ENDFOR
 \STATE $z{(i)} = h^{(L)}{(i)}$
\end{algorithmic}
\end{algorithm}

\noindent \textbf{Combine Function.}
In the neighborhood aggregation phase, we describe the representation of a node in terms of its neighbors. Moreover, to retain the knowledge of each node's original features, we propose to use the combine function. Combine function concatenates the aggregated embeddings of the neighbors from the current layer with the self-embedding of the node from the previous layer and is given by:
\begin{equation}
\label{comb}
h^{(l)}{(i)} = \sigma{\Bigg(W^{(l)}\bigg(h^{(l-1)}{(i)} \mathbin\Vert h^{(l)}{\Big(N(i)\Big)}\bigg) \Bigg)}
\end{equation}

\noindent where $h^{(l-1)}{(i)}$ represents embedding of node $i$ from layer $(l-1)$ and $h^{(l)}{(N(i))}$ represents aggregated embedding of the neighbors of node $i$. $W^{(l)}$ is the trainable parameters, $\mathbin\Vert$ denotes the concatenation operator, and $\sigma$ represents the non-linearity ReLU. \\

\noindent \textbf{High Order Propagation.}
GraphSHS stacks multiple layers (Neighborhood Aggregation phase and Combine Function phase) to capture information from the $l$-hop neighbors of a node. The output of layer $(l-1)$ acts as an input for layer $l$, whereas the embeddings at layer $0$ are initialized with the initial features of the nodes. Stacking $l$ layers will recursively formulate the embeddings $h^{(l)}(i)$ for node $i$ at the end of $l^{th}$ layer as:
\begin{equation}
z{(i)} = h^{(l)}{(i)}, \,\,\,\, \forall i \in V
\end{equation}

\noindent where $z(i)$ denotes the final embedding of node $i$ at the end of $l^{th}$ layer ($l$ = $1,...,L$). For the purpose of node classification, we pass the final embeddings $z(i)$ of all the nodes through the Softmax Layer. The softmax layer maps the embeddings of the nodes to the probabilities of two classes, i.e., SHS and normal node. The model is then supervised to learn to differentiate between SHS and normal nodes using the labelled data available.\\

\noindent {\textit{\textbf{Algorithm 1:}} The algorithm begins by initializing node embeddings based on their initial input features, as described in Line 1. It then proceeds to iterate through multiple layers, as indicated in Lines 2-7. Within these layers, the algorithm consistently refines the node embeddings by capturing feature information from neighboring nodes (Line 4) and incorporating the node's self-embedding (Line 5). This iterative process ensures that the embeddings evolve as the depth increases, enabling them to capture more complex relationships and structural information within the graph. Finally, upon completing all the iterations, the algorithm produces the final node embeddings in Line 8, which encode the inherent structure and characteristics of the graph.}

\subsubsection{Model Training} 
In order to differentiate between SHSs and normal nodes, we train GraphSHS using \textit{Binary Cross-Entropy Loss} with the actual labels known for a set of nodes. The loss function $\mathcal{L}$ is computed as:
\begin{equation}
\label{eq_loss}
\mathcal{L}(\theta) = {-}\frac{1}{t}{\sum_{i=1}^{t}\bigg(y(i)\log{\hat{y}(i)} + (1-y(i)) \log{(1-\hat{y}(i))\bigg)}}
\end{equation}
\noindent where $y$ is the actual label of a node and $\hat{y}$ is the label predicted by GraphSHS, $t$ is the number of nodes in the training data for which the labels are known, and $\theta$ are the set of model parameters.

\begin{theorem}[\textbf{Loukas \protect{\citep{loukas2019graph}}}] \label{thm1}
\textbf{A simple message passing architecture of GraphSHS is sufficient to solve the SHSs discovery problem if it satisfies the following conditions: each node in the graph is distinctively identified; functions (Neighborhood aggregation and Combine function) computed within each layer $l$ are Turing-Complete; the architecture is deep enough, and the width is unbounded.}
\end{theorem}

Here, depth indicates the number of layers in the architecture and width is the number of hidden units. Simple message passing graph neural networks are proven to be \textit{universal} if the four conditions mentioned above are satisfied \citep{loukas2019graph}. Therefore, we adopt a simple message passing graph neural network architecture to solve the SHSs discovery problem, and our architecture satisfies these conditions. We believe that the universal characteristic of graph neural networks enables our model to discover SHS nodes with high accuracy. This argument is confirmed by our experimental results, as reported in Section 5. Notably, we choose not to include the unique identifiers (node ids) in our node features as SHSs are {\em equivariant} to node permutation. In other words, we can interpret our graph neural network GraphSHS as a function that maps a graph with $n$ nodes to an output vector of size $n$, where the $i^{th}$ coordinate of the output specifies whether node $i$ is a SHS or not. Since any permutation on the graph nodes would also permute the output exactly in the same way, and thus, what our model is trying to learn is an \textit{equivariant function} (by definition). Keriven et al. \citep{keriven2019universal} proposed a simple graph neural network architecture that does not require unique identifiers and shows that the network is a universal approximator of equivariant functions.
It should be noted that the theoretical results of Keriven et al. \citep{keriven2019universal} do not apply directly to message-passing graph neural networks that are more often used in practice. We do not have proof that unique identifiers are not necessary for our model, as we are using message-passing graph neural networks. We do not include unique identifiers as a design choice.

\begin{figure}[h!]
 \centering
 \includegraphics[width=0.9\columnwidth]{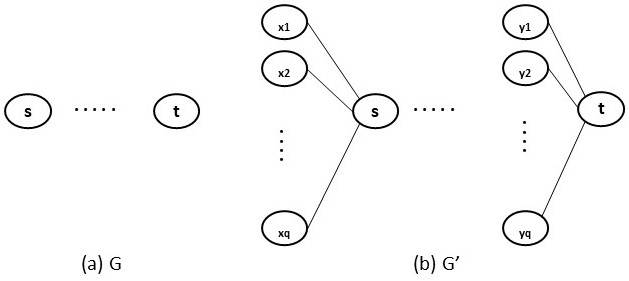}
 \caption{$G'$ is constructed from $G$.}
 \label{fig:hard}
\end{figure}

\begin{theorem} \label{thm2}
\textbf{To calculate the SHSs discovery problem (discovering high betweenness centrality nodes), the depth of GraphSHS (with constant width) should be at least $\Omega(\sqrt{n}/\log n)$.}
\end{theorem}

\begin{proof} Let $G=(V,E)$ be an instance of shortest $s$-$t$ path problem \citep{loukas2019graph} in an undirected graph with source node $s$, destination node $t$ and $|V|=n$, as shown in Figure \ref{fig:hard}(a). The shortest $s$-$t$ path problem aims to find the nodes that lie on the shortest path from node $s$ to $t$. We construct an instance of discovering high betweenness centrality nodes (SHSs) problem in another undirected graph $G'$ from $G$, as illustrated in Figure \ref{fig:hard}(b). 
We add a set of nodes $X = \{x1, x2, ...,xq\}$ which are connected to node $s$ via undirected edges $\{x1\mathdash s, x2\mathdash s, ..., xq\mathdash s\}$. Similarly, we add another set of nodes $Y = \{y1, y2, ...,yq\}$  which are connected to node $t$ via undirected edges $\{y1\mathdash t, y2\mathdash t, ..., yq \mathdash t\}$. Our goal is to discover high betweenness centrality nodes (SHSs) in graph $G'$.

Let us assume that the value of $q$ is $cn$, where $c$ is a constant $\geq 3$ . For computation, we assume $q=3n$; then, for every node that lies on the shortest $s$-$t$ path, there are $9n^2$ shortest paths that go through these nodes. For the rest of the nodes that do not lie on the shortest $s$-$t$ path, the shortest paths in $G'$ that go through these nodes are:
\begin{enumerate}[noitemsep,topsep=0pt]
    \item The shortest paths between the nodes of the original graph $G$ in $G'$. For this case, there are at most $n^2$ shortest paths passing through the nodes that do not lie on the shortest $s$-$t$ path in $G'$.
    \item The shortest paths between the nodes of set $X$ to the nodes of the original graph $G$ in $G'$. For this case, there are at most $(3n \times n)$, i.e., $3n^2$ shortest paths passing through the nodes that do not lie on the shortest $s$-$t$ path.
    \item The shortest paths between the nodes of set $Y$ to the nodes of the original graph $G$ in $G'$. There are at most $(3n \times n)$, i.e., $3n^2$ shortest paths passing through the nodes that do not lie on the shortest $s$-$t$ path.
\end{enumerate}

There are $9n^2$ shortest paths going through the nodes that lie on the shortest $s$-$t$ path, which is greater than the total number of shortest paths, i.e., at most $(n^2 + 3n^2 + 3n^2)$ going through the nodes that do not lie on the shortest $s$-$t$ path, i.e.,  $(n^2 + 3n^2 + 3n^2 \leq 9n^2)$. According to the definition of betweenness centrality, a node would have a high betweenness centrality if it appears on many shortest paths. Our analysis shows that more number of shortest paths go through those nodes that lie on the shortest $s$-$t$ path; therefore, high betweenness centrality nodes must also lie on the shortest $s$-$t$ path. 

In this way, if we can find the high betweenness centrality nodes (SHSs) in the graph, then we can solve the shortest $s$-$t$ path problem. Corollary 4.3 of \citep{loukas2019graph} already showed that for approximating (to a constant factor) the shortest $s$-$t$ path problem, a message-passing graph neural network must have a depth that is at least $\Omega(\sqrt{n}/\log n)$ assuming constant model width. Hence, this depth lower bound also applies to our SHSs discovery problem.
\end{proof}

\begin{figure*}[t!]
 \centering
 \includegraphics[width=0.73\paperwidth]{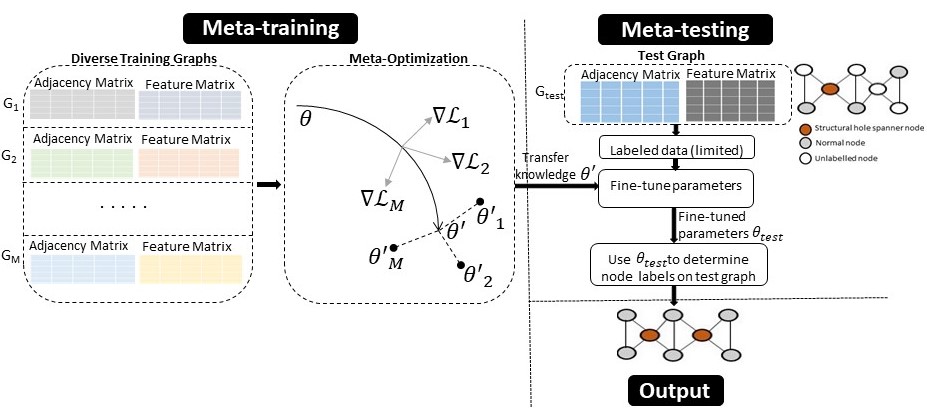}
 \caption{The overall architecture of the proposed model Meta-GraphSHS. The left side illustrates the meta-training phase that outputs a generalized parameter $\theta'$, which is transferred to the meta-testing phase. Meta-testing phase utilizes $\theta'$ as a good initialization point and fine tune $\theta'$ using labelled data from the test graph to obtain $\theta_{test}$, which can be used to obtain labels of unlabeled nodes in the test graph.}
 \label{fig:meta-gnn}
\end{figure*}

{\subsection{Complexity Analysis}
\textit{\textbf{Training time.}} To train the GraphSHS model on a network of 5000 nodes, the convergence time is around 15 minutes, which includes the time to compute the ground truth labels and features of the nodes for the training graph. Notably, we train the model only once and then utilize the trained model to predict the nodes' labels for any input graph.}

{\textit{\textbf{Inference complexity.}} In the application step of GraphSHS, we apply the trained GraphSHS model to a given network for discovering SHSs. To determine the labels of the nodes, the model computes embeddings for each node. Algorithm 1 shows that computing the nodes' embedding takes $\mathcal{O}(LnN)$ time, where $L$ is the depth (number of layers) of the network, $n$ is the number of nodes, and $N$ is the average number of node neighbors. In practice, adjacency matrix multiplication is used for Line 3-6 in Algorithm 1, and if the graph is densely connected, then the complexity for Line 3-6 is $\mathcal{O}(n^2)$. Theoretically, we showed that the lower bound on depth $L$ is $\Omega(\sqrt{n}/\log n)$; therefore, the \textit{theoretical lower-bound complexity} for application step of GraphSHS is $\mathcal{O}(n^2 \sqrt{n}/\log n)$. On the other hand, we experimentally showed that the depth $L$ of the GraphSHS is a small constant $(L = 4)$, and most of the real-world networks are sparse; therefore, the practical time
complexity for the application step of GraphSHS turns out to be
$O(m)$, i.e., linear in the number of edges.}

\subsection{Meta-GraphSHS: Discovering SHSs in Diverse Networks}
In this section, we discuss our proposed meta-learning based model \textbf{\textit{Meta-GraphSHS}} that aims to discover SHS nodes across diverse networks. The crucial challenge in this task is to capture the inter-graph differences and customize the model according to the new diverse graph (test graph). Meta-GraphSHS discovers SHSs in the new test graph $G_{test}$ (called meta-testing graph) by training the model on a set of diverse training graphs $G_{train}$ = $\{G_1, G_2,...,G_M\}$ (called meta-training graphs). The distribution over graphs is considered as a distribution over tasks and we consider the training task corresponding to each training graph in $G_{train}$ as $\tau$ = $\{\tau_1, \tau_2,.., \tau_M\}$. Similarly, $\tau_{test}$ is the task corresponding to test graph $G_{test}$. We further refer to the training and testing node set in all tasks $\tau$ as support set $S$ and query set $Q$. Let $S_i$ represent the support set, and $Q_i$ represent the query set for $G_i$, and $f_{\theta}$ represent the model, where $\theta$ is a set of model parameters.

Meta-GraphSHS addresses above mentioned challenge by first learning the general parameters from diverse training graphs $G_{train}$ and utilizing these parameters as a good initialization point for the test graph $G_{test}$. The learned general parameters are fine-tuned using the small number of available labelled data of the test graph\footnote{Fine-tune aims to precisely adjust the learned general model parameters in order to fit with the test graph.} (support set of $\tau_{test}$) and the obtained updated parameters are used to determine the labels of unlabeled nodes in the test graph (query set of $\tau_{test}$). In this way, Meta-GraphSHS avoids the need for repeated model training on each type of different graph (which is a time-consuming task) by designing a customized model that can be quickly adapted to the test graph under consideration in a few gradient steps, given only a few labelled nodes in the test graph. Figure \ref{fig:meta-gnn} illustrates the overall architecture of the proposed model Meta-GraphSHS. Our goal is to reach an \textit{``almost trained model"} that quickly adapts to the new graph. The performance of the Meta-GraphSHS is determined via meta-testing on the testing task $\tau_{test}$, by fine-tuning the model on the support set of $\tau_{test}$ and evaluating on the query set of $\tau_{test}$. Meta-GraphSHS uses Model-Agnostic Meta-Learning for updating the gradients during training {\citep{finn2017model}}. {Model-Agnostic Meta-Learning (MAML) is a machine learning technique that enables models to quickly adapt to new tasks with minimal training data. It works by learning an initial set of model parameters that, when fine-tuned with a small dataset for a specific task, allows the model to generalize and perform well on that task.} The procedure for meta-training and meta-testing are discussed below:\\

\noindent \textbf{Meta-training.}
During training, we intend to learn a set of generalizable parameters that act as a good initialization point for Meta-GraphSHS with the aim that the model rapidly adapts to the new task (test graph $G_{test}$) within a few gradient steps. For $M$ learning tasks $\{G_1, G_2,...,G_M\}$, we first adapt the model's initial parameters to every learning task individually. We use $\mathcal{L}_{\tau_i}(\theta)$ to represent the loss function for task $\tau_i$. We utilize the same procedure as that of GraphSHS to train the model on task $\tau_i$ and compute the loss function for the same using Equation \ref{eq_loss}. After computing the loss, updated model parameter $\theta_{i}'$ are computed using gradient descent. We update the parameters as follows:
\begin{equation}
\label{12}
\theta_{i}' = \theta - \alpha \nabla_\theta \mathcal{L}_{\tau_i}(\theta)
\end{equation}

\noindent where $\alpha$ represents learning rate and $\theta$ becomes $\theta_{i}'$ when adapting to the task $\tau_i$. We just describe 1 gradient step in Equation \ref{12}, considering many gradient steps as a simple extension \citep{finn2017model}.
Since there are $M$ learning tasks, $M$ different variants of the initial model are constructed (i.e., $f_{\theta_1'}, ...,f_{\theta_M'}$). We train the model parameters by optimizing the performance of $f_{\theta_i'}$ on all tasks. Precisely, the meta-objective is given by:
\begin{equation}
\label{35}
\theta = \mathop{\arg \min}\limits_{\theta} \sum_{i=1}^{M} \mathcal{L}_{\tau_i}(\theta_{i}')
\end{equation}

Notably, optimization is performed over $\theta$, and the objective function is calculated using the updated parameters ${\theta_i'}$. The model parameters are optimized in such a way that only a few gradient steps are needed to adjust to the new task, maximizing the model's prediction performance on the new task. We use stochastic gradient descent to perform optimization across tasks. The parameter $\theta$ is updated as below:
\begin{equation}
\label{45}
\theta \leftarrow \theta - \gamma\nabla_\theta \sum_{i=1}^{M}\mathcal{L}_{\tau_{i}}(\theta_{i}') 
\end{equation}

\noindent where $\gamma$ represents meta-learning rate. The learned general parameter is then transferred to the meta-testing phase. The training procedure for Meta-GraphSHS is presented in Algorithm \ref{meta-algo}.\\

\noindent {\textbf{\textit{Algorithm \ref{meta-algo}:}} The algorithm begins by initializing the model's parameters, represented as $\theta$, with random values (Line 1). It then iterates until an early stopping condition is met. In this loop, the algorithm traverses each of the training graphs, denoted as $G_1$ through $G_M$ (Line 3). For each graph, $G_i$, the algorithm performs the following tasks: it splits the nodes into two sets, the support set $S_i$ and the query set $Q_i$ (Line 4). Following this, the algorithm computes the loss function and updates the model parameter $\theta$ using gradient descent (Lines 5 and 6), and these steps are repeated for each graph in the training dataset (Line 7). Subsequently, the algorithm proceeds to Line 8, where the model's parameters are updated using Equation 9 based on the remaining nodes in the graphs (query set), denoted as $Q_1$ through $Q_M$. Finally, after completing the training loop, Line 10 describes the fine-tuning phase, where the model's parameters, $\theta$, are fine-tuned on the testing graph, $G_{test}$, using a specified loss function.}\\

\noindent \textbf{Meta-testing.} The model in meta-testing phase is initialized with the learned parameters from meta-training phase, due to which the model is already almost trained. We then feed the support set $S_{test}$ of test graph $G_{test}$ as input to the model and fine-tune the learned model parameters precisely to fit with $G_{test}$. Since the model is already almost trained, it just takes a few gradient steps to fine-tune the model. After fine-tuning, the model performance is assessed on query set $Q_{test}$ of test graph $G_{test}$.

\begin{algorithm}[t!]
\caption{Training procedure for Meta-GraphSHS}
 \label{meta-algo}
 \begin{algorithmic}[1]
 \renewcommand{\algorithmicrequire}{\textbf{Input:}}
 \renewcommand{\algorithmicensure}{\textbf{Output:}}
 \REQUIRE $G_{train}$=$\{G_1, G_2,...,G_M\}$ and $G_{test}$
 \ENSURE Parameters: $\theta$
 \STATE Initialize $\theta$ randomly
 \WHILE{not early-stop}
 \FOR{$G_i$ = $G_1, G_2,...,G_M$}
 \STATE Split labelled nodes of $G_i$ into $S_i$ and $Q_i$
 \STATE Evaluate $\nabla_\theta \mathcal{L}_{\tau_i}(\theta)$ for $S_i$ 
 \STATE Compute parameter $\theta_{i}'$ using Equation \ref{12}
 \ENDFOR
 \STATE Update $\theta$ on $\{Q_1, Q_2,...,Q_M\}$ using Equation \ref{45}
 \ENDWHILE
 \STATE Fine-tune $\theta$ on $G_{test}$ using loss function
\end{algorithmic}
\end{algorithm}

\section{EXPERIMENTS}
We discuss the performance of the proposed models GraphSHS and Meta-GraphSHS by performing exhaustive experiments on widely used datasets. We first discuss the experimental setup. We then report the performance of GraphSHS on various synthetic and real-world datasets, followed by the performance of Meta-GraphSHS. Lastly, we present the parameter sensitivity analysis and application improvement.

\begin{table*}[t!] 
\caption{Summary of synthetic datasets.}
\label{summary_syn}
\renewcommand{\arraystretch}{1.1}
\centering 
\begin{tabular}{lll}\hlineB{1.5} 
\textbf{Graph type} & \multicolumn{2}{c}{\textbf{Graph generating parameters}} \\ \hlineB{1.5}
\multirow{2}{*}{\textbf{Erdos-Renyi Graphs}} & Number of nodes & \specialcell[t]{5000, 10000, 20000, 50000} \\ \cline{2-3}
 &{Probability of adding a random edge} & 0.001\\\cline{2-3}
 
 & Number of nodes & \specialcell[t]{{100000, 150000}} \\ \cline{2-3}
 &{Probability of adding a random edge} & 0.0001\\\hlineB{1.5}
\multirow{4}{*}{\textbf{Scale-Free Graphs}} & Number of nodes & \specialcell[t]{5000, 10000, 20000, 50000, {100000, 150000}}\\ \cline{2-3}
 & Alpha & 0.4 \\\cline{2-3}
 & Beta & 0.05 \\\cline{2-3}
 & Gamma & 0.55 \\ \hlineB{1.5} 
\end{tabular} 
\end{table*}

\begin{table}[t!] 
\caption{Summary of real-world datasets.}
\label{summary_real}
\renewcommand{\arraystretch}{1.1}
\centering 
\begin{tabular}{llll}\hlineB{1.5} 

\textbf{Dataset} & \textbf{Nodes} & \textbf{Edges} & \textbf{Avg degree} \\ \hlineB{1.5}
ca-CondMat & 21,363 & 91342 & 8.55 \\
email-Enron & 33,696 & 180,811 & 10.73 \\
coauthor & 53,442 & 255,936 & 4.8 \\
com-DBLP & 317,080 & 1,049,866 & 6.62 \\
com-Amazon & 334,863 & 925,872 & 5.53 \\ \hlineB{1.5}
\end{tabular} 
\end{table}

\subsection{Experimental Setup for GraphSHS}

\subsubsection{\textit{Datasets}} 
We report the effectiveness and efficiency of GraphSHS on various datasets. The details of synthetic and real-world datasets are discussed below.

\noindent \textbf{\textit{\\Synthetic Datasets}}. Considering the features of the Python NetworkX library, we used this library to create two types of synthetic graphs, namely Erdos-Renyi graphs (ER) \citep{erdHos1959random} and Scale-Free graphs (SF) \citep{onnela2007structure}. For each type, we generate test graphs of six different scales: $5000$, $10000$, $20000$, $50000$, {$100000$ and $150000$} nodes by keeping the parameter settings the same. In addition to these test graphs, we generate two graphs of 5000 nodes, one of each type (ER and SF) for training GraphSHS. Notably, for each type of graph (ER and SF), we train the model on a graph of 5000 nodes and test the model on all scales of graphs ($5000$, $10000$, $20000$, $50000$, $100000$ and $150000$ nodes). Table \ref{summary_syn} presents the summary of graph generating parameters for synthetic datasets\footnote{Due to the computational challenges in computing ground truths for large-scale graphs, we limit the maximum number of edges to 200000 for ER and SF graphs with 100000 and 150000 nodes.}.\\

\noindent \textbf{\textit{Real-World Datasets}}. We use five real-world datasets to determine GraphSHS performance. Table \ref{summary_real} presents the summary of these datasets, and the details are discussed below:
\begin{itemize}[leftmargin=*]
\item \textbf{{ca-CondMat \citep{leskovec2007graph}}} is a scientific collaboration network from arXiv. This network covers collaborations between the authors who have submitted papers in condensed matter category.
\item \textbf{{email-Enron \citep{leskovec2009community}}} is a communication network of emails where nodes denote the addresses, and edge connects two nodes if they have communicated via email. 
\item \textbf{{coauthor \citep{lou2013mining}}} is an author-coauthor relationship network. It consists of coauthor relationships obtained from papers published in major computer science conferences. 
\item \textbf{{com-DBLP \citep{yang2012defining}}} is a coauthor network. Nodes represent the authors, and an edge connects the authors if they have published at least one paper together.
\item \textbf{{com-Amazon \citep{yang2012defining}}} is a customer-product network obtained from amazon website. Nodes represent the customers, and edge connects the customers who have purchased the same product.
\end{itemize}

\subsubsection{\textit{{Evaluation Metrics}}} 
For baselines and GraphSHS, we measure the effectiveness and efficiency in terms of accuracy and running time, respectively.\\

\noindent {\textbf{{Accuracy.}} Accuracy is defined as the proportion of correct predictions, made by an approach, to the total number of predictions made across all classes.}

{\begin{equation*}
\text{Accuracy} = \frac{\text{Number of Correct Predictions}}{\text{Total Number of Predictions}}
\end{equation*}}

\noindent {\textbf{Run time.} Run time is defined as the amount of time it takes for an approach to discover SHS nodes on a given dataset.}}

\subsubsection{\textit{Baselines}} 
We compare GraphSHS with the two representative SHS identification algorithms:

\begin{itemize}[leftmargin=*]
\item \textbf{{Constraint.}} Constraint is a heuristic solution to discover SHSs in the network \citep{burt1992structural}. It measures the degree of redundancy among the neighbors of the node. Constraint $C$ of a node $i$ is defined as:
\begin{equation*}
\begin{aligned}
C(i) = \sum_{j\in N(i)}{\left({p_{ij}}+\sum_{q}{p_{iq} p_{qj}}\right)}^2, \quad q\neq i,j
\end{aligned}
\end{equation*}
where $N(i)$ is neighbors of node $i$, $q$ is the node in the ego network other than node $i$ and $j$, and $p_{ij}$ represents the weight of edge $(i,j)$.

\item \textbf{{Closeness Centrality.}} The closeness centrality of a node is the reciprocal of sum of length of the shortest paths from the node to all other nodes in the graph  \citep{bavelas1950communication}. Rezvani et al. \citep{rezvani2015identifying} used closeness centrality as a base to propose an algorithm Inverse Closeness Centrality (ICC), for discovering SHSs in the network. Closeness Centrality (CC) of node $i$ is calculated as:
\begin{equation*}
\begin{aligned}
CC(i) = \frac{1}{\sum_{j \in V}\text{SP}(i,j)}
\end{aligned}
\end{equation*}
where $\text{SP}(i,j)$ is the shortest path between node $i$ and $j$.

\item {\textbf{Vote Rank Algorithm.} Vote Rank is an iterative algorithm to identify top-$k$ decentralized spreaders with the best spreading ability. This algorithm uses a voting scheme to rank nodes in a graph, where each node votes for its in-neighbors, and the node with the highest number of votes is selected in each iteration \citep{zhang2016identifying}.}

\end{itemize}


\begin{table*}[h!] 
\caption{Classification accuracy (\%) on synthetic datasets of different scales. Top-$5$\%, Top-$10$\% and Top-$20$\% indicate the percentage of nodes labelled as SHSs. Bold results indicate the best results among the proposed and all baselines.}
\label{result_syn}
\renewcommand{\arraystretch}{1.1}
\centering 
\begin{tabular}{p{1.7cm}p{3.5cm}p{1.2cm}p{1.2cm}p{1.2cm}p{1.2cm}p{1.2cm}p{1.2cm}p{1.2cm}p{1.2cm}p{1.2cm}p{0.9 cm}} \hlineB{1.4} 
\textbf{Scale $\downarrow$} & \multirow{4}{*}{\textbf{Method}} & \multicolumn{2}{c}{\textbf{{Top-$5$\%}}} & \multicolumn{2}{c}{\textbf{{Top-$10$\%}}}& \multicolumn{2}{c}{\textbf{{Top-$20$\%}}} \\ \cmidrule(lr){3-4} \cmidrule(lr){5-6} \cmidrule(lr){7-8}
\textbf{Dataset $\rightarrow$} & & \textbf{SF} & \textbf{ER} & \textbf{SF} & \textbf{ER} & \textbf{SF} & \textbf{ER} \\ \hlineB{1.5}
 & Constraint & 94.23 & 93.98 & 91.57 & 91.05 & {88.26} & {84.35}\\ 
5,000 & Closeness centrality & 94.58 & 93.38 & 88.66 & 90.7 & 86.54 & 82.26\\
& {Vote Rank} & {96.12} & {95.43} & {92.41} & {91.13} & {\textbf{92.79}} & {\textbf{85.51}}\\
 & GraphSHS (Proposed) & \textbf{96.78} & \textbf{95.66} & \textbf{93.66} & \textbf{91.30} & 87.65 & 83.22\\ \hlineB{1.5}
 & Constraint & 94.02 & 94.45 & 90.87 & 92.02 & {87.24} & {85.64} \\
10,000& Closeness centrality & 94.81 & 94.09 & 89.21 & {92.34} & 85.75 & 80.75\\
 & {Vote Rank} & {95.29} & {95.18} & {92.04} & {\textbf{93.82}} & {\textbf{94.41}} & {\textbf{{87.98}}} \\
 & GraphSHS (Proposed) & \textbf{96.44} & \textbf{95.29} & \textbf{93.23} & 90.89 & 86.92& 82.45 \\ \hlineB{1.5}
 & Constraint & 95.02 & 93.97& 88.23 & 90.61 & {88.32} & \textbf{87.41} \\ 
20,000 & Closeness centrality & 94.29 & 94.35 & 87.71 & \textbf{{91.39}} & 82.78 & 80.34\\
& {Vote Rank} & {95.01} & {94.88} & {92.59} & {91.25} & {\textbf{89.64}} & {{87.28}} \\
 & GraphSHS (Proposed) & \textbf{96.31} & \textbf{95.23} & \textbf{92.97} & 90.56 & 85.80 & 81.22 \\\hlineB{1.5}

 & Constraint & 94.93 & 93.85 & 87.12 & 88.65 & 84.77 & \textbf{82.36}\\
50,000 & Closeness centrality & 93.95 & 91.89 & 85.27 & 84.91 & 81.60 & 72.37 \\
& {Vote Rank} & {94.64} & {93.27} & {91.54} & {87.83} & {85.49} &  {81.22}\\
 & GraphSHS (Proposed) & \textbf{95.03} & \textbf{94.81} & \textbf{92.01} & \textbf{89.49} & \textbf{85.55} & 80.24 \\ \hlineB{1.5} \colorrows{} 
& Constraint & NA & 90.49 & NA & 82.08 & NA  & 68.15\\
100,000 & Closeness centrality & 93.51 & 87.18 & 88.06 & 85.48 & 84.20 & \textbf{85.33}  \\
& Vote Rank & 94.18 & 92.72 & 91.43 & 86.76 & {87.11} & 80.73\\
 & GraphSHS (Proposed) &  \textbf{94.93} & \textbf{93.75} & \textbf{91.84} & \textbf{87.9} & \textbf{88.37} & 80.6 \\ \hlineB{1.5}
 
& Constraint & NA & 89.40 & NA & 82.56 & NA & 68.03\\
150,000 & Closeness centrality & 93.04 & 91.61&  91.14  &  88.50 &  {\textbf{{89.03}}} & \textbf{86.74} \\
& Vote Rank & 93.92 & 91.93 & 90.73 & 86.92 & 85.87 & 78.17\\
 & GraphSHS (Proposed) & \textbf{94.25} & \textbf{93.56} & \textbf{91.69} & \textbf{89.35} & {88.82}  & 83.42 \\ 
 \hlineB{1.5}
\end{tabular} 
\end{table*}

\subsubsection{\textit{Ground Truth Computation}} \label{ground}
For all the datasets under consideration, we used the Python library NetworkX to calculate nodes' SHS score (BC). Besides, for large scale graphs, i.e., com-DBLP and com-Amazon, we used the SHS score (BC) reported by AlGhamdi et al. \citep{alghamdi2017benchmark}. {The authors performed parallel implementation of the Brandes algorithm, utilizing 96,000 CPU cores on a supercomputer to compute exact BC values for large graphs {\citep{alghamdi2017benchmark}}. We were not able to perform experiments on very large synthetic networks, as it is} {computationally challenging to compute the ground truth BC for larger graphs using normal system configurations; therefore, we limit the synthetic network size to 150000 nodes.} 

After computing the SHS score of the nodes, we sort the nodes in descending order of their score values. We label the high score $k$\% nodes as SHS nodes and the rest as normal ones. We evaluate the performance of GraphSHS for three different values of $k$, i.e., $5$, $10$ and $20$. Labelled graphs are used to train GraphSHS, and we assess the performance of GraphSHS on the test graphs.

\subsubsection{\textit{Training Details}}
We perform all the experiments on a Windows 10 PC with a CPU of 3.20 GHz and 16 GB RAM. We implement the code in PyTorch. We fix the number of layers to 4 and the embedding dimension to 128. Parameters are trained using Adam optimizer with a learning rate of 0.01 and weight decay $5e-4$. We train the GraphSHS for 200 epochs on ER graph of 5000 nodes and evaluate the performance on test ER graphs of all scales. We adopted the same training and testing procedure for SF graphs. Since real-world networks demonstrate attributes similar to SF graphs; therefore, we train our model on an SF graph of 5000 nodes and test the model on real-world datasets. Besides, we used an inductive setting where test graphs are invisible to the model during the training phase.

\begin{table*}[t!] 
\caption{Run time (sec) comparison of different algorithms on synthetic datasets of different scales. Bold results indicate the best results, and second best results are underlined.}
\label{time_syn}
\renewcommand{\arraystretch}{1.1}
\centering 
\begin{tabular}{p{1.5cm}p{1.5cm}p{1.9cm}p{3cm}p{1.9cm}p{1.9cm}p{1.2cm}} \hlineB{1.5} 
\textbf{Scale} & \textbf{Dataset} & \textbf{Constraint} & \textbf{Closeness centrality } & {\textbf{Vote Rank}} & \textbf{GraphSHS (Proposed)}& \textbf{Speedup} \\ \hlineB{1.5}
\multirow{2}{*}{5,000} & SF & 16013.2 & {40.1} & {\underline{17.18}} & \textbf{0.09} & 190.9x \\ 
 & ER & \underline{5.8} & 45.9 & {32.26} &\textbf{0.1} & 58x\\\hline
\multirow{2}{*}{10,000} & SF & 21475.3 & {199.4} & {\underline{29.72}} & \textbf{0.3} & 99.1x\\
 & ER & \underline{67.2} & 286.1 & {316.88} & \textbf{0.5} & 134.4x\\\hline
\multirow{2}{*}{20,000} & SF & 24965.3 & {836.7} & {\underline{155.89}} & \textbf{0.7} & 222.7x \\
 & ER & \underline{884.7} & 1820.7 &  {2970.47} & \textbf{1.75} & 505.5x\\\hline
\multirow{2}{*}{50,000} & SF & 28336.1 & {5675.8} &  {\underline{1088.39}} & \textbf{2.5} & 435.3x\\
 & ER & 27754.2 & \underline{2055.4} & {3987.62} & \textbf{12.6} & 163.1x\\\hlineB{1.5}
\multirow{2}{*}{100,000} & SF & NA & 4442.9 &  {\underline{2164.68}} & \textbf{15.4} & 140.5x\\
 & ER & 29345.1 & 13746.9 &  {\underline{3512.17}} & \textbf{27.4} & 128.2x \\\hlineB{1.5}
 \multirow{2}{*}{150,000} & SF & NA & {3143.1} & {\underline{1592.46}} & \textbf{21.6} & 73.7x\\
 & ER & 31601.73 &  22338.3 &  {\underline{15631.67}} & \textbf{33.7} & 463.8x \\\hlineB{1.5}
 
\end{tabular} 
\end{table*}

\subsection{Performance of GraphSHS on Synthetic Datasets}
Tables \ref{result_syn} and \ref{time_syn} report the accuracy and run time of the comparative algorithms and GraphSHS on synthetic graphs. Table \ref{result_syn} shows that GraphSHS achieves higher classification accuracy than the baselines. For example, in the SF graph of 5000 nodes, GraphSHS performs better than the baselines, closeness centrality, constraint and vote rank by achieving Top-$5$\% accuracy of 96.78\%, whereas the best accuracy achieved by the baseline is 96.12\%. Besides, GraphSHS is $190.9$ times faster than the best result for the same scale and type of graph, as reported in Table \ref{time_syn}. For the ER graph of 10000 nodes, although GraphSHS sacrifices 2.93\% in Top-$10$\% accuracy in contrast to the top accuracy (vote rank); however, it is over $134$ times faster. 
{For a large-scale SF graph of 100000 nodes, GraphSHS achieves higher accuracy than the baselines by achieving a Top-$5$\% accuracy of 94.93\%, whereas the best accuracy achieved by the baseline is 94.18\% (vote rank). Notably, the constraint algorithm cannot complete the computation for the SF graph of 100000 and 150000 nodes within three days, so we put NA corresponding to its accuracy and time in the results. Moreover, our proposed model achieves the best accuracy for SF and ER graphs of 150000 nodes in the case of Top-$5$\% and Top-$10$\% accuracy; however, for Top-$20$\%, closeness centrality achieves higher accuracy.} To avoid unfair comparison, we have not considered the training time of the model as none of the baseline algorithms needs to be trained. Hence, it is logical not to count the training time. Moreover, GraphSHS converges rapidly, and the convergence time is around 15 minutes. In addition, our model works in multi-stages. We can train the model whenever we have time and later use it for discovering SHSs. However, all the baselines identify SHSs in one stage only.

Table \ref{time_syn} reports the running time of baselines and GraphSHS. For a small scale ER graph of 5000 nodes, GraphSHS takes 1 sec to discover SHSs, whereas closeness centrality takes 45.9 sec and vote rank takes 32.26 sec. For a large-scale ER graph of 50000 nodes, GraphSHS takes less than 13 sec to discover SHSs. However, constraint, vote rank and closeness centrality require a large amount of time to discover SHSs in large-scale networks. For ER graph of 50000 nodes, constraint took around 7.5 hours, whereas both closeness centrality and vote rank took around 1 hour to discover SHSs. {For ER graph of 150000 nodes, GraphSHS takes less than 34 sec to discover SHSs. However, all the baselines require a large amount of time to discover SHSs and GraphSHS is 463.8 times faster than the most efficient baseline.} The results prove that our model has a considerable efficiency advantage over other models in run time.

The proposed model GraphSHS consistently achieves the best Top-$5$\% accuracy for ER as well as SF graphs of all scales. GraphSHS achieves the highest Top-$10$\% accuracy for  most of the cases; however, vote rank achieves better Top-$10$\% accuracy for ER graphs of 10000 nodes and closeness centrality for ER graphs of 20000 nodes. The vote rank algorithm outperforms most of the comparative methods for ER and SF graphs in terms of Top-$20$\% accuracy. Although other algorithms achieve better accuracy than GraphSHS in a few cases, but our model runs faster. GraphSHS is at least 58 times faster than the baselines on synthetic graphs. Results from Table \ref{result_syn} show that the classification accuracy is inversely proportional to the size of the network. In addition, there is a decrease in Top-$k$\% accuracy as we increase the value of $k$. 

Table \ref{type_syn} presents the generalization accuracy of the proposed model GraphSHS across different types of graphs. We train the GraphSHS on ER and SF graphs separately, and test on both types of graphs. For this analysis, we only consider graphs of 5,000 nodes for training and testing. The results demonstrate that GraphSHS attains the best accuracy when the training graph is similar to testing graphs.

\begin{table}[t!] 
\caption{Generalization accuracy (\%) of GraphSHS on different types of synthetic datasets.}
\label{type_syn}
\renewcommand{\arraystretch}{1.1}
\centering 
\begin{tabular}{p{3.2cm}|p{1.9cm}p{1.9cm}} \hlineB{1.5} 

\backslashbox{\textbf{Train} $\downarrow$}{\textbf{Accuracy}}{\textbf{Test$\rightarrow$}}
&\makebox[5em]{\textbf{ER\_5,000}}&\makebox[5em]{\textbf{SF\_5,000}}\\\hlineB{1.5}

ER\_5,000 & \textbf{95.66} & 94.22 \\
SF\_5,000 & 93.16 & \textbf{96.78} \\\hlineB{1.5}
\end{tabular} 
\end{table}

\begin{table*}[h!] 
\caption{Classification accuracy (\%) on real-world datasets. Top-$5$\%, Top-$10$\% and Top-$20$\% indicate the percentage of nodes labelled as SHSs. Bold results indicate the best results among the proposed and all baselines.}
\label{result_real}
\renewcommand{\arraystretch}{1.1}
\centering 
\begin{tabular}{p{2cm}p{3.5cm}p{2cm}p{2cm}p{1.7cm}} \hlineB{1.5} 
\textbf{Dataset}& \textbf{Method} & {\textbf{{Top-$5$\%}}} & {\textbf{{Top-$10$\%}}}& {\textbf{{Top-$20$\%}}} \\ \hlineB{1.5}

 & Constraint & 94.41 & 90.18 & {86.23}\\
ca-CondMat & Closeness centrality & 95.05 & 89.78 & 82.56\\
& {Vote Rank} & {95.59} & {90.15} & {\textbf{89.44}} \\
 & GraphSHS (Proposed) & \textbf{95.73} & \textbf{90.43} & 83.23\\ \hline
 
 & Constraint & 95.77 & 91.87 & \textbf{87.38}\\ 
email-Enron & Closeness centrality & 95.41 & 90.98 & 83.71 \\ 
& {Vote Rank} & {95.93} & {93.01} & {87.28}\\
 & GraphSHS (Proposed) & \textbf{96.2} & \textbf{93.13} & 86.49 \\ \hline
 & Constraint & 93.60 & 90.77 & \textbf{86.61}\\
coauthor & Closeness centrality & 94.4 & 88.95 & 81.07\\
& {Vote Rank} &  {94.59} &  {\textbf{93.6}} & {86.53}\\
 & GraphSHS (Proposed) & \textbf{95.03} & {91.28} & 80.91 \\ \hline
 
 & Constraint & 92.4 & \textbf{91.42} & \textbf{84.21}\\ 
com-DBLP & Closeness centrality & \textbf{95.1} & 89.9 & 80.2\\ 
& {Vote Rank} &{NA}  &{NA}  &{NA}\\
 & GraphSHS (Proposed) & 93.11 & 89.2 & 81.24 \\ \hline
 
 & Constraint & 94.61 & \textbf{88.12} & \textbf{83.15}\\
com-Amazon & Closeness centrality & 93.13 & 87.30 & 77.83 \\ 
& {Vote Rank} &{NA}  &{NA}  &{NA} \\
 & GraphSHS (Proposed) & \textbf{94.71} & 85.21 & 78.23\\ \hlineB{1.5}
 
\end{tabular} 
\end{table*}

\begin{table*}[h!] 
\caption{Run time (sec) comparison of different algorithms on real-world datasets. Bold results indicate the best results and the second best results are underlined.}
\label{time_real}
\renewcommand{\arraystretch}{1.1}
\centering 
\begin{tabular}{p{2cm}p{2cm}p{3cm}p{2cm}p{2cm}p{1.3cm}} \hlineB{1.5} 
\textbf{Dataset} & \textbf{Constraint} & \textbf{Closeness centrality } & {\textbf{Vote Rank}} &  \textbf{GraphSHS (Proposed)}& \textbf{Speedup} \\ \hlineB{1.5}
ca-CondMat & {1403.2} & 2853.4 & {\underline{983.88}} & \textbf{1.07} & 919.5x\\
email-Enron & \underline{1968.5} & 2903.4 & {2541.6} & \textbf{2.2} & 894.7x\\
coauthor & \underline{417.8} & 5149.6 &  {3948.53} & \textbf{2.5} & 167.1x\\
com-DBLP & \underline{8574.9} & 38522.1 &{NA} & \textbf{19.2} & 446.6x\\
com-Amazon & \underline{4533.4} & 42116.9 &{NA} & \textbf{18.9} & 239.8x\\
\hlineB{1.5}
\end{tabular} 
\end{table*}

\subsection{Performance of GraphSHS on Real-World Datasets}

This section evaluates GraphSHS performance on five real-world datasets. Since real-world networks exhibit some characteristics similar to that of SF graphs; therefore, we train our model on an SF graph (SF graph of 5000 nodes having the same properties as discussed in Table \ref{summary_syn}) and test the model on real-world datasets. We present the Top-$k$\% accuracy and running time of the baselines in Tables \ref{result_real} and \ref{time_real}, respectively. The results illustrate that GraphSHS attains competitive Top-$k$\% accuracy compared to other baselines. Nevertheless, considering the trade-off between accuracy and run time, GraphSHS runs much faster than the baselines. Take the example of the ca-CondMat network; GraphSHS performs better than the baselines by achieving Top-$5$\% accuracy of 95.73\% and Top-$10$\% accuracy of 90.43\%. 
Although vote rank performs better in the Top-$20$\% accuracy for the same network; however, GraphSHS is 919.5 times faster. {In the email-Enron graph, GraphSHS performs better than the baselines by achieving the highest Top-$5$\% and Top-$10$\% accuracy.} On the other hand, if we take an example of a large-scale network, such as com-Amazon, constraint outperforms GraphSHS in Top-$10$\% and Top-$20$\% accuracy; however, GraphSHS is 239.8 times faster than the best baseline. {The vote rank algorithm cannot complete the computation for com-DBLP and com-Amazon networks within 3 days, probably due to the large network size (approximately 1,000,000 edges). Therefore, we have included NA corresponding to its accuracy and time in the results.} GraphSHS achieves the best Top-$5$\% accuracy in four real-world networks and the best Top-$10$\% accuracy in three out of five networks. However, for the Top-$20$\% accuracy, constraint algorithm is more accurate. Table \ref{time_real} shows that GraphSHS achieves a minimum speedup of 167.1 and is up to 919.5 times faster than the baseline algorithms. The run time comparison indicates the efficiency advantage of our model over other baselines. Furthermore, our results proved that the proposed simple graph neural network architecture GraphSHS is sufficient to solve the SHSs discovering problem on real-world networks.

\begin{table*}[h!] 
\caption{Summary of dataset for evaluating Meta-GraphSHS.}
\label{meta_data}
\renewcommand{\arraystretch}{1.4}
\centering 
\begin{tabular}{c|c|c|c} \hlineB{1.5} 

\textbf{Dataset} & \textbf{Types of subgraph} & \textbf{\#Subgraphs} & \textbf{\#Nodes in each subgraph} \\ \hlineB{1.5}
\multirow{3}{*}{{Synthetic graph}} & Erdos-Renyi graphs & \multirow{3}{*}{{36}} & \\ \cline{2-2}
 & Scale-Free graphs & & 1000 to 5000 \\ \cline{2-2}
 & Gaussian Random Partition graphs & & \\
\hlineB{1.5}

\multirow{2}{*}{{Real-world graph}} & ca-CondMat graphs & \multirow{2}{*}{{24}} & 1000 to 3000\\ \cline{2-2}
 & email-Enron graphs & &  \\ \cline{2-2}
\hlineB{1.5}

\end{tabular} 
\end{table*}

{
\begin{table}[h!] 
\caption{Classification accuracy (\%) of Meta-GraphSHS.}
\label{result_meta}
\renewcommand{\arraystretch}{1.4}
\centering 
\begin{tabular}{c|c|c} \hlineB{1.5} 
\textbf{Dataset}& \textbf{Method} & \textbf{Accuracy} \\ \hlineB{1.5}
\multirow{5}{*}{Synthetic graph} & \multirow{1}{*}{{Constraint}} & {88.2} \\ \cline{2-3}
& \multirow{1}{*}{{Closeness centrality}} & {88.4} \\ \cline{2-3}
& \multirow{1}{*}{{Vote Rank}} & {90.8} \\ \cline{2-3}

 & \multirow{1}{*}{GraphSHS} & 93.5 \\ \cline{2-3}
 & \multirow{1}{*}{Meta-GraphSHS} & 96.2 \\ \hlineB{1.5} 

 \multirow{5}{*}{Real-world graph} & \multirow{1}{*}{{Constraint}} & {87.7} \\ \cline{2-3}
& \multirow{1}{*}{{Closeness centrality}} & {89.5} \\ \cline{2-3}
& \multirow{1}{*}{{Vote Rank}} & {89.3} \\ \cline{2-3} 
 & \multirow{1}{*}{GraphSHS} & 92.1 \\ \cline{2-3}
 & \multirow{1}{*}{Meta-GraphSHS} & 94.8 \\ \hlineB{1.5} 
\end{tabular} 
\end{table}}

\subsection{Performance of Meta-GraphSHS}
In order to obtain a classifier Meta-GraphSHS that can discover SHSs in diverse networks, we train our model on different types of networks. {We evaluate the performance of our model on the following synthetic and real-world datasets. The summary of dataset is presented in Table \ref{meta_data}.}

\begin{itemize}[leftmargin=*]
\item {\textbf{Synthetic graph.} We generate one synthetic graph consisting of 36 sub-graphs of 3 different types, i.e., Erdos-Renyi, Scale-Free, and Gaussian Random Partition graphs. The graph contains 12 sub-graphs of each type, and each sub-graph consists of a minimum of 1000 nodes and a maximum of 5000 nodes. }

\item {\textbf{Real-world graph.} We obtain one real-world graph by combining 2 diverse real-world graphs, i.e., ca-CondMat and email-Enron (refer Table \ref{summary_real} for properties of these graphs). For each of these graphs, we disconnect the original graph to obtain 12 much smaller subgraphs. In this way, the overall graph contains 24 sub-graphs (12 of each type) and each sub-graph consists of a minimum of 1000 nodes and a maximum of 3000 nodes. }
\end{itemize}

We follow the procedure discussed in Section \ref{ground} for obtaining the ground truths for the graphs and label the top 5\% nodes in each of the sub-graph as SHS nodes. We use 80\% of the sub-graphs for training (meta-training), and 20\% for testing (meta-testing). We train Meta-GraphSHS for 200 epochs, and set $\alpha$ to 0.1 and $\gamma$ to 0.001. The training sub-graphs are used to optimize the model parameters (to learn generalizable parameters by observing multiple graphs from different domains). Only 50\% of the nodes in the testing sub-graphs are labelled. The labelled nodes in testing sub-graphs are used to fine-tune the trained model to accurately determine labels for the rest of the nodes in the test graphs. {Table \ref{result_meta} shows the accuracy achieved by Meta-GraphSHS for discovering SHSs in diverse synthetic and real-world graphs compared to that of baselines. For diverse synthetic graphs, Meta-GraphSHS discovers SHS nodes with high accuracy of 93.5\% and outperforms GraphSHS by an accuracy of 2.7\%. {Vote Rank achieves an accuracy of 90.8\%, which is significantly lower than Meta-GraphSHS.} 

{For diverse real-world graphs, results demonstrate that Constraint attains an accuracy of 87.7\%. Additionally, Closeness Centrality yields an accuracy of 89.5\%, while Vote rank achieves an accuracy of 89.3\%. Notably, GraphSHS method exhibits a performance level of 92.1\%. However, Meta-GraphSHS model surpasses all the baselines by achieving a significantly higher accuracy rate of 94.8\%, highlighting its superior performance.} Our previous results from Table \ref{result_syn} and Table \ref{result_real} illustrate that even though GraphSHS discovers SHSs with high accuracy when trained and tested on graphs from the same domain; however, the accuracy decreases when GraphSHS is tested on graphs from different domains than what the model is trained on. The reason for the low accuracy of GraphSHS in the case of diverse graphs is that the model is not able to capture the inter-graph differences. {The performance of Meta-GraphSHS on both synthetic and real-world graphs shows that machine learning models explicitly designed for a particular task outperform the models designed for generalized tasks. The advantage of meta learning models is that they learn from experience and quickly adapt to new tasks with minimal training data. This allows them to achieve high accuracy, even with diverse data. This is why once trained, Meta-GraphSHS generalizes well and discovers SHSs from diverse networks with high accuracy.}

\subsection{Parameter Sensitivity}
We perform experiments on the real and synthetic networks to determine the impact of parameters on the accuracy of both the proposed models, GraphSHS and Meta-GraphSHS. Particularly, we study the sensitivity of the number of layers (depth) and embedding dimensions for the models. We vary the number of layers and embedding dimension among \{1, 2, 3, 4, 5, 6\} and \{16, 32, 64, 128, 256\}, respectively. 
Figures \ref{fig:real} and \ref{fig:syn} show the parameter sensitivity of GraphSHS on real-world and synthetic datasets, respectively. The results illustrate that the accuracy is relatively low for fewer aggregation layers (depth), as shown in Figures \ref{fig:real}(a) and \ref{fig:syn}(a). The reason for low accuracy is insufficient aggregated information due to the limited reachability of the nodes. Our results show that initially, the SHSs identification accuracy increases with the increase in the number of layers (model depth); however, if we increase the depth of the model over four layers, the accuracy starts decreasing. The reason for this is the over-smoothing problem  \citep{li2018deeper, yang2020toward, pasa2021multiresolution}. Besides, results from Figures \ref{fig:real}(b) and \ref{fig:syn}(b) show that for higher embedding dimensions, GraphSHS performs better as higher embedding dimensions provide the GraphSHS with more ability to represent the network.

{Our reasoning behind the improved accuracy of GraphSHS, with the increase in the number of layers and embedding dimensions, is further supported by the parameter sensitivity analysis results of the Meta-GraphSHS model. The parameter sensitivity analysis of Meta-GraphSHS on synthetic and real-world datasets is demonstrated in Figure \ref{fig:metaablagtion}. The results indicate that the accuracy is comparatively low for a smaller number of aggregation layers, as depicted in Figure \ref{fig:metaablagtion}(a), and the accuracy increases as the number of layers increases. However, the accuracy begins to decline after four layers. Similarly, Figure \ref{fig:metaablagtion}(b) presents the accuracy of Meta-GraphSHS on varying the number of embedding dimensions. Firstly, the accuracy increases with the increase in embedding dimensions; however, the accuracy starts deteriorating on increasing the embedding dimension to 256 or higher. The reason for the deteriorating performance of Meta-GraphSHS on increasing the embedding dimensions beyond 128 is the similar node embeddings that make it difficult for the model to distinguish between the nodes and, consequently, the model mislabels the nodes.}

\begin{figure*}[h!]
 \centering
 \includegraphics[width=0.8\paperwidth]{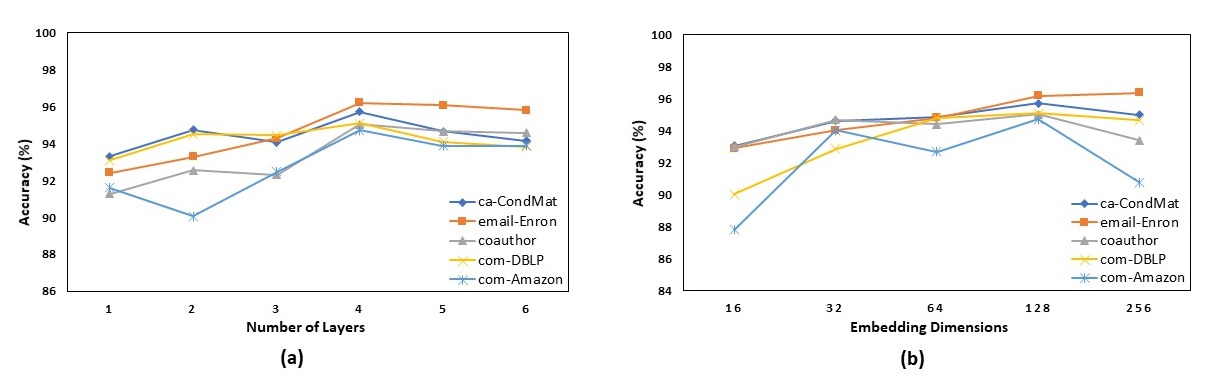}
 \caption{Parameter sensitivity of GraphSHS on real-world datasets.}
 \label{fig:real}
\end{figure*}

\begin{figure*}[h!]
 \centering
 \includegraphics[width=0.8\paperwidth]{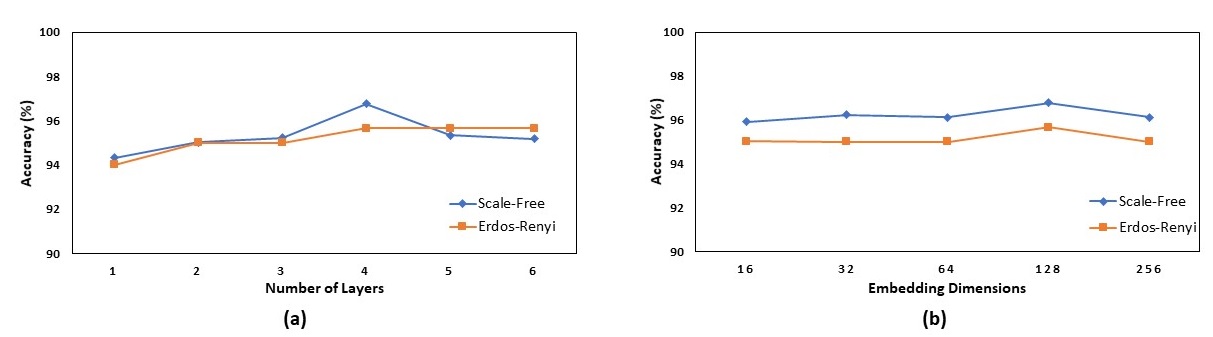}
 \caption{Parameter sensitivity of GraphSHS on synthetic datasets.}
 \label{fig:syn}
\end{figure*}

\begin{figure*}[h!]
 \centering
 \includegraphics[width=0.78\paperwidth]{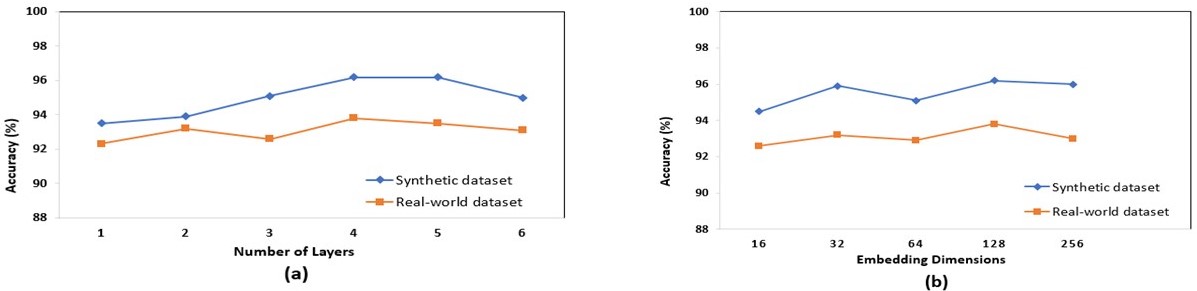}
 \caption{{Parameter sensitivity of Meta-GraphSHS on synthetic and real-world datasets.}}
 \label{fig:metaablagtion}
\end{figure*}

\subsection{Application Improvement}
GraphSHS can be used to discover SHSs in a dynamic network, where nodes and edges change over time. For example, on Facebook and Twitter, links appear/disappear whenever a user friend/unfriend others on Facebook or follow/unfollow others on Twitter. As a result, discovered SHSs change, and hence, it is essential to track the new SHSs in the updated network. Traditional algorithms are highly time-consuming and might not work efficiently for dynamic networks. Additionally, it is highly possible that the network has already been changed by the time these algorithms re-compute SHSs. Therefore, we need a fast heuristic that can quickly update SHSs in dynamic networks.

We can use our proposed model GraphSHS for discovering SHSs in dynamic networks. Even if training the model takes a few hours to learn the evolving pattern of the network, we only need to train the model once, and after that, whenever there is a change in the network, our trained GraphSHS can identify the new SHSs within a few seconds. We compare our proposed GraphSHS with the solution designed by Goel et al. \citep{goel2021maintenance} that discovers SHSs in dynamic networks. 
We start with an entire network and arbitrarily delete 100 edges, one edge at a time, and calculate the average speedup of GraphSHS over dynamic solution \citep{goel2021maintenance}. As shown in Table \ref{appl}, GraphSHS is at least $89.8$ times faster than \citep{goel2021maintenance}. This confirms the efficiency of the proposed model in dynamic networks. 
\begin{table}[h!] 
\caption{Performance of GraphSHS on dynamic networks.}
\label{appl}
\renewcommand{\arraystretch}{1.4}
\centering 
\begin{tabular}{p{1.3cm}|p{1.3cm}|c|p{1.3cm}} \hlineB{1.5} 
\textbf{Dataset} & \textbf{\# Nodes} & \textbf{\# SHSs discovered} & \textbf{Speedup} \\ \hlineB{1.5}
\multirow{3}{*}{{Scale-Free}} & \multirow{3}{*}{{5,000}} & 1 & 89.8x \\ \cline{3-4}
 & & 5 & 139.5x \\ \cline{3-4}
 & & 10 & 163.7x \\ 
\hlineB{1.5}
\end{tabular} 
\end{table}

\subsection{Discussion}
Our experiments on various datasets show that our simple message-passing graph neural network models are sufficient to solve the SHSs discovering problem. Our proposed model GraphSHS provides a significant run time advantage over other algorithms. GraphSHS is at least 167.1 times faster than the baselines on real-world networks and at least 58 times faster than the baselines on synthetic networks. Even though we trained GraphSHS on the synthetic SF graph, the model achieved high accuracy when tested on real-world graphs. This shows the inductive nature of our proposed model, where the model can be trained on one graph and used to predict SHSs in another graph. Besides, once trained, Meta-GraphSHS generalizes well and identifies SHSs from diverse networks with high accuracy of 96.2\% for synthetic graphs and 94.8\% for real-world graphs. The following observations are the potential reasons behind the success of the proposed models:
\begin{enumerate}[leftmargin=*,noitemsep,topsep=0pt]
    \item Our proposed graph neural network based models follow a similar architecture to that of message-passing graph neural networks, which are proved to be universal under sufficient conditions \citep{loukas2019graph}. Our model meets those conditions, and we believe that the \textit{universal characteristics} of the message passing graph neural network enable our model to capture the relevant features that are important for discovering SHSs, which is confirmed from our experimental results.
    \item We train the proposed model in an end-to-end manner with the exact betweenness centrality values as ground truth. Similar to successful applications of deep learning on text or speech,  the model generally learns and performs well if provided with sufficient training data.
\end{enumerate}
In addition, theoretically, we showed that the depth of the GraphSHS should be at least $\Omega(\sqrt{n}/\log n)$. However, practically, deep GNNs suffer from the over-smoothing issue that leads to the embeddings of nodes indistinguishable from each other. We conduct parameter sensitivity analysis to investigate the effect of model depth (number of layers) on the accuracy of discovering SHSs in the network. Our experimental results showed that after a few number of layers, the performance of GraphSHS starts deteriorating. The reason for the dropping performance of GraphSHS is similar node embeddings, which results in the model being unable to differentiate between the nodes and, hence, mislabel them. Therefore, in our experiments, we made the necessary adjustment and used four layers instead in order to avoid the over-smoothing problem.

\section{CONCLUSION}

Structural hole spanner identification problem has various real-world applications such as information diffusion, community detection etc. However, there are two challenges that need to be addressed 1) to discover SHSs efficiently in large scale networks; 2) to discover SHSs effectively across diverse networks. This paper investigated the power of message-passing GNNs for identifying SHSs in large scale networks and diverse networks. We first transformed the SHS identification problem into a learning problem and designed an efficient message-passing GNN-based model GraphSHS that identifies SHS nodes in large scale networks with high accuracy. We then proposed another effective meta-learning model Meta-GraphSHS that discovers SHSs across different types of networks. Meta-GraphSHS learns general transferable knowledge during the training process and then quickly adapts by fine-tuning the model parameters for each new unseen graph. We used an inductive setting that enables the proposed models to be generalizable to new unseen graphs. Theoretically, we showed that the proposed graph neural network model needs to be at least $\Omega(\sqrt{n}/\log n)$ deep to calculate the SHSs discovery problem. To evaluate the model's performance, we performed empirical analysis on various datasets. Our experimental results demonstrated that the proposed models achieved high accuracy. GraphSHS is at least 167.1 times faster than the baselines on large scale real-world networks, showing a considerable advantage in run time over the baseline algorithms.

In our future work, we will design effective graph neural network models for discovering SHSs in dynamic networks. Specifically, we will focus on developing methods for efficiently updating the node embedding whenever there is any change in the network.

\bibliographystyle{elsarticle-harv}
\bibliography{Refer}
\end{document}